\documentclass[aps,prl,notitlepage,amsmath,amssymb,amsfonts,superscriptaddress,nofootinbib,twocolumn]{revtex4-2}

\makeatletter
\def\@hangfrom@section#1#2#3{\normalsize\@hangfrom{#1#2}#3}%\MakeTextUppercase{#3}}%
\def\@hangfroms@section#1#2{\normalsize#1#2}%\MakeTextUppercase{#2}}%
\makeatother

\usepackage{graphicx,soul}% Include figure files
\usepackage{dcolumn}% Align table columns on decimal point
\usepackage{bm}% bold math
\usepackage{bbm} % bb 1

\usepackage{dsfont}
\usepackage{enumerate}

\usepackage[dvipsnames, x11names]{xcolor}
\usepackage[colorlinks=true]{hyperref}
\hypersetup{
    colorlinks=true,
    citecolor=blue,
    linkcolor=blue,
    urlcolor=NavyBlue,
    anchorcolor=Purple
}

\usepackage[percent]{overpic}
\usepackage{tabularx}
\usepackage{multirow}
\usepackage{CJK}
\usepackage{orcidlink}

\usepackage{physics}
\usepackage{braket}

\newcommand{\bluefour}[1]{\textcolor{Blue4}{#1}}

\def\be{\begin{equation}}
\def\ee{\end{equation}}
\def\bea{\begin{eqnarray}}
\def\eea{\end{eqnarray}}

\begin{document}

\begin{CJK*}{UTF8}{}

    \title{Quantum oscillations in proximity to high-angular-momentum band inversion}

    \author{Jiahao~Yang~(\CJKfamily{gbsn}杨家豪) \orcidlink{0000-0001-7670-2218}}
    \affiliation{International Center for Quantum Materials, School of Physics, Peking University, 
    Beijing 100871, China}
    \affiliation{Beijing Key Laboratory of Quantum Devices, Peking University, Beijing 100871, China}

    \author{Gang~v.~Chen~(\CJKfamily{gbsn}陈钢) \orcidlink{0000-0001-9339-6398}} 
    \email{chenxray@pku.edu.cn}
    \affiliation{International Center for Quantum Materials, School of Physics, Peking University, 
    Beijing 100871, China}
    \affiliation{Beijing Key Laboratory of Quantum Devices, Peking University, Beijing 100871, China}
    \affiliation{Collaborative Innovation Center of Quantum Matter, 100871, Beijing, China}

    \begin{abstract}
    Quantum oscillations provide a fundamental 
    probe of electronic structure in magnetic fields, revealing the Fermi surface topology 
    in metals, and/or the quasiparticle properties even in the insulating regimes.  
    Here we study quantum oscillations in minimal models of high-angular-momentum (HAM) band inversion for both a chiral two-band model and a time-reversal-invariant four-band model. 
    In the former case, finite oscillations can appear at the hybridized Chern-insulating regime due to thermal-activated excitations.
    In the latter case, interference between the two time-reversal-related blocks drives a strong deviation from the Lifshitz-Kosevich form, 
    producing a non-monotonic temperature dependence of the oscillation amplitude 
    and characteristic suppression temperatures whose number follows the angular momentum $l$. 
    These results identify experimentally accessible signatures of HAM band inversion and provide a framework for other discrete-symmetry-related hybridizations and excitonic pairings.
    \end{abstract}

    \date{\today}
    \maketitle

\end{CJK*}

% \tableofcontents 

% \section{Introduction}

\noindent\bluefour{\it Introduction.}---Quantum oscillations in electronic transport 
and thermodynamic properties, conventionally described by the Lifshitz-Kosevich (LK) theory, 
serve as a powerful probe of Fermi surface topology 
in metals~\cite{shoenberg1984MagneticOscillationsMetals}.
Modern interests have invoked the quantum oscillations in the
weakly gapped semiconductors and topological band insulators, 
where these oscillations acquire an additional layer of richness~\cite{knolle2015QuantumOscillationsFermi,mikitik2004BerryPhase,zhang2005ExperimentalObservation,xiang2018QuantumOscillations}.  
The Berry phase accumulated along the cyclotron orbits modifies 
the phase offset of the oscillations, enabling the direct experimental detection 
of non-trivial band topology~\cite{mikitik2004BerryPhase,zhang2005ExperimentalObservation,zhang2016QuantumOscillationNarrowGap}. 
Beyond the single-particle band structures, quantum oscillations have emerged 
as a sensitive tool for exploring systems with emergent gauge fields 
and fractionalized quasiparticles. For instance, in quantum spin liquids 
and other correlated insulators, the coupling of external magnetic field 
to internal gauge fluxes can generate the oscillatory 
low-energy density-of-states (LEDOS) responses that defy the conventional LK behavior, 
offering potential signatures of 
emergent gauge-matter structures~\cite{knolle2015QuantumOscillationsFermi,knolle2017AnomalousHaas,
allocca2024FluctuationdominatedQuantumOscillations, thiagarajan2025NatureTopological,motrunich2006OrbitalMagnetic,
sodemann2018QuantumOscillationsa,chowdhury2018MixedvalenceInsulators}.

\begin{figure}[b]
    \centering
    \begin{overpic}[width=0.7\columnwidth]{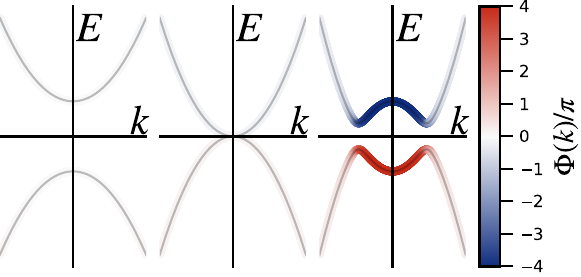}
    \put(2,45){(a)}
    \put(28,45){(b)}
    \put(55,45){(c)}
    \end{overpic}
    \begin{overpic}[width=0.27\columnwidth]{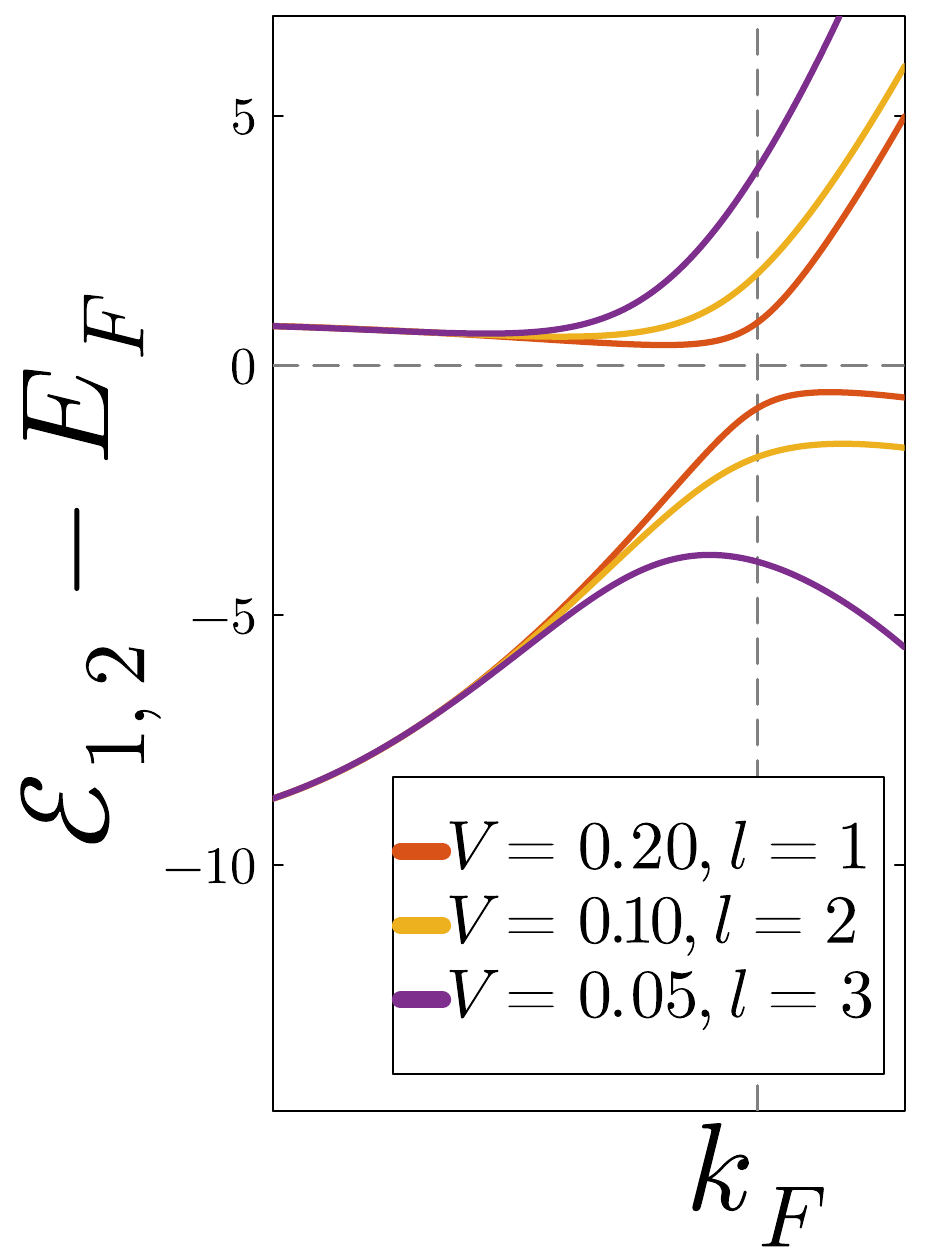}
        \put(0,90){(d)}
    \end{overpic}
    \caption{HAM band inversion. 
    (a) Trivial band insulator with ${\delta\mu<0}$.
    (b) Quadratic band touching at ${\delta\mu=0}$. 
    (c) Chern insulator with ${\delta\mu>0}$. Here, Chern number ${C=2}$ for ${l=2}$,
    and the Berry phase $\Phi(k)$ is plotted with ${\Phi(k_F)=2\pi}$. 
    The same structure extends to ${l\geq 2}$, where the inverted insulator 
    has the Chern number ${C=l}$.  
    (d) Hybridized dispersions $\mathcal{E}_{1,2}(k)$ for ${l=1,2,3}$ 
    with different hybridizations $V$.  
    ``$k_F$'' denotes the crossing point of the unhybridized bands 
    and is measured in inverse-length units. }
    \label{fig:BandInversion}
\end{figure}

The concept of high-angular-momentum (HAM) band inversion (${l \geq 2}$) 
provides a useful framework for understanding an important class of unconventional quantum-oscillation phenomena.
This inversion, protected by the crystalline rotational symmetry,    
describes the reversal of two bands with relative angular momentum $l$ 
and drives a transition from a trivial insulator to a topological insulator  
or Chern insulator (see Fig.~\ref{fig:BandInversion})~\cite{venderbos2018HigherAngularMomentum}. 
It naturally appears in a variety of platforms, 
moir\'{e} graphene systems~\cite{navarro-labastida2025TopologicalPhasea}, 
multilayer rhombohedral graphene~\cite{zhang2015SpontaneousChiral,dong2024StabilityAnomalous,davenport2026BerryCurvature}, 
    three-dimensional pyrochlore iridates~\cite{moon2013NonFermiLiquidTopological,yao2018LuttingerSemimetal}, 
and even the spinon spectrum of the Kitaev honeycomb model~\cite{thiagarajan2025NatureTopological, zhang2022TheoryKitaevModel,xu2012SpinLiquid}. 
In the latter fractionalized context, the interplay between the external 
magnetic field and the internal emergent gauge field fundamentally 
alters the nature of quantum oscillations, as the fractionalized quasiparticles 
couple to both fields directly or indirectly. 
While such rich physics warrants future investigation, 
a prerequisite is a thorough understanding of quantum oscillations 
arising from the HAM band inversion itself, 
without additional complications from gauge field or strong correlations. 
The minimal models of HAM band inversions provide a clean and accessible route 
to studying the LEDOS oscillations under a magnetic field. 
In this Letter, we focus on the quantum oscillations arising from purely electronic HAM band inversion. 
This problem can be understood in terms of two classes of minimal models.

The first class consists of two-band models, where the conduction 
and valence bands are coupled by a chiral hybridization $(k_x \pm i k_y)^{l}$ 
with ${l \ge 2}$~\cite{venderbos2018HigherAngularMomentum}. 
These models encompass all orbital-based constructions, 
including for example the $s$-$d$ and $s$-$f$ systems on various lattices. 
In the inverted regime, such a two-band description yields a Chern insulator 
with Chern number ${C = l}$. The second class extends to time-reversal-invariant systems 
by combining two copies of the two-band model with opposite chirality, 
following the Bernevig-Hughes-Zhang (BHZ) scheme~\cite{bernevig2006QuantumSpin}. 
This four-band model describes a transition from a normal insulator to a topological insulator 
protected by rotation and inversion symmetry, characterized by $l$ helical edge modes. 
While both classes share the quadratic band structure and enhanced correlation 
at the inversion criticality,  their different symmetries and band topology  
lead to fundamentally distinct oscillation behaviors under magnetic  fields~\cite{venderbos2018HigherAngularMomentum,hu2018FractionalExcitonicInsulator,gassner2025FluxAttachmentTheory}. 
In particular, we show that the chiral two-band model supports finite quantum oscillations even in the insulating regime, arising from the thermal population of the hybridized bands. 
By contrast, in the four-band case the oscillation amplitude displays a nonmonotonic temperature dependence 
that departs from the Lifshitz-Kosevich form, with characteristic temperatures whose number tracks $l$ due to interference between the two time-reversal-related blocks.
We therefore analyze them in turn, first for a chiral two-band model and then for the corresponding time-reversal-invariant four-band model.

% \section{Model}

\noindent\bluefour{\it Two-band model.}---We begin with a minimal
chiral two-band model consisting of a light $d$ band and a heavy $f$ band 
with annihilation operators $d_{\bm{k}}$ and $f_{\bm{k}}$, respectively. 
Here $d$ and $f$ refer to any two bands that are involved in the HAM band inversion. 
The two-dimensional model reads 
$\mathcal{H} = \sum_{\bm{k}} 
\begin{pmatrix}
    d_{\bm{k}}^\dagger & 
    f_{\bm{k}}^\dagger
\end{pmatrix}
\mathcal{H}_{\bm{k}} 
\begin{pmatrix}
    d_{\bm{k}} \\
    f_{\bm{k}}
\end{pmatrix}$
with
\begin{align} 
\label{eq:Model}
    \mathcal{H}_{\bm{k}} & = 
    \begin{pmatrix}
        \epsilon_d(\bm{k})  & V(k_x - i k_y)^{l} \\
        V(k_x + i k_y)^{l} & \epsilon_f(\bm{k})
    \end{pmatrix},
\end{align}
where $V$ is the chiral hybridization strength, 
and $l$ is the relative angular momentum between the two bands.
$\epsilon_{\lambda}(\bm{k})$ (${\lambda \equiv d,f}$) is the unhybridized dispersion  
with ${\epsilon_{\lambda}(\bm{k})=\hbar^2k^2/2m_{\lambda} - \mu_{\lambda}}$ 
with $m_{\lambda}$ the effective mass. ${\delta\mu \equiv \mu_d - \mu_f}$
defines the detuning parameter for the band inversion, 
such that the system lies in the normal phase for ${\delta\mu < 0}$ 
and in the inverted phase for ${\delta\mu > 0}$, 
with the topological critical point at ${\delta\mu = 0}$ 
(see Fig.~\ref{fig:BandInversion}). For ${\delta\mu > 0}$, 
the two unhybridized bands cross at the Fermi wavevector 
${k_F = \sqrt{2m_1 \delta\mu}/\hbar}$ and Fermi energy 
${E_F = \delta\mu m_1 /m_d}$, where ${m_{1,2} \equiv m_dm_f/(m_f\pm m_d)}$.
The two-band Hamiltonian~\eqref{eq:Model} generates the spectrum
$\mathcal{E}_{1,2}(k)=\frac{1}{2}
\big(\epsilon_d(k)+\epsilon_f(k) \big)
\pm \frac{1}{2} \sqrt{[\epsilon_d(k)-\epsilon_f(k)]^2+4V^2k^{2l}}$.
In Fig.~\ref{fig:BandInversion}(d), hybridized bands 
exhibit a direct band gap at $k_F$, $\Delta_D=2V(2m_1\delta\mu)^{l/2}/\hbar^l$, 
and an indirect gap for finite-momentum excitations, $\Delta_I=2[\sqrt{m_dm_f}/(m_d+m_f)]\Delta_D$.
Moreover, a smaller hybridization $V$ suffices to open the gap at larger $l$.

\begin{figure}[t]
    \begin{overpic}[width=0.32\linewidth]{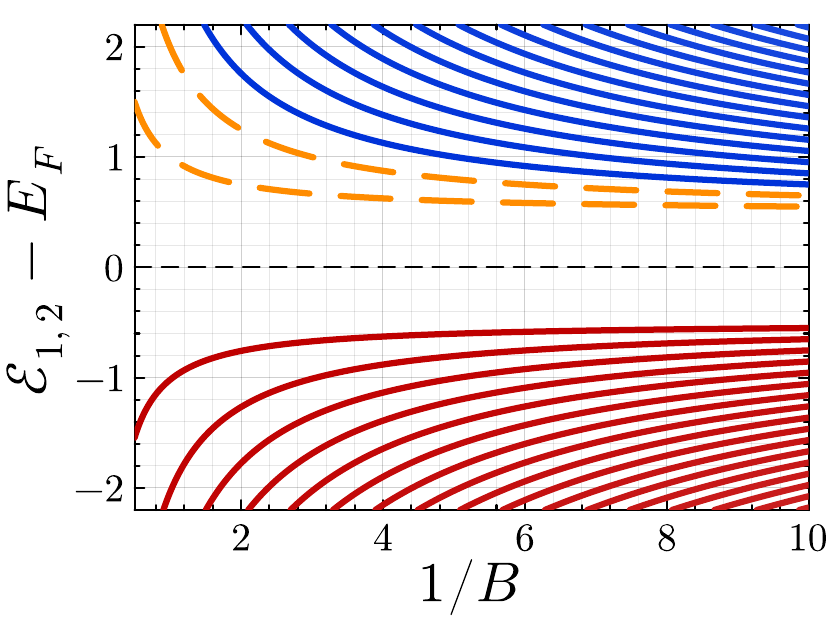}
        \put(0,70){(a)} % 在左上角添加标号
    \end{overpic}
    \begin{overpic}[width=0.32\linewidth]{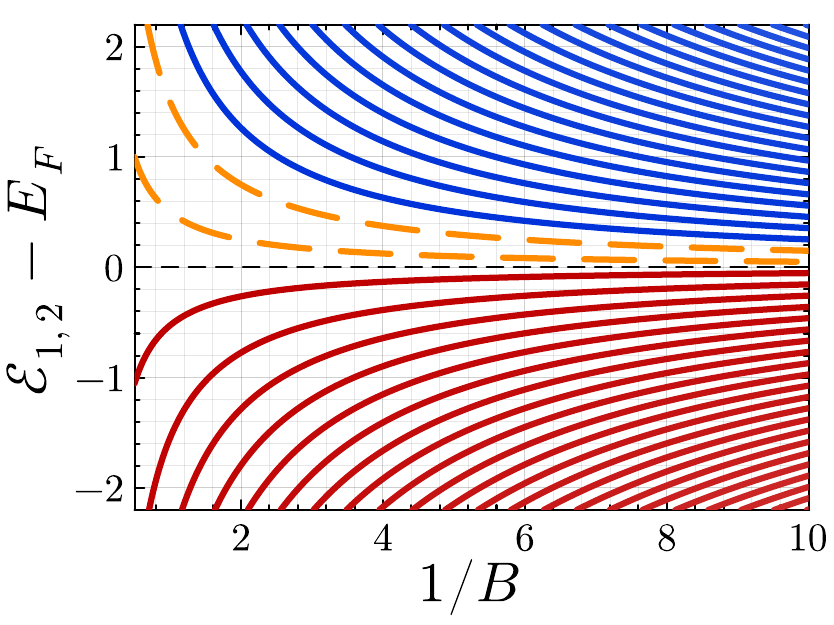}
        \put(0,70){(b)} % 在左上角添加标号
    \end{overpic}
    \begin{overpic}[width=0.32\linewidth]{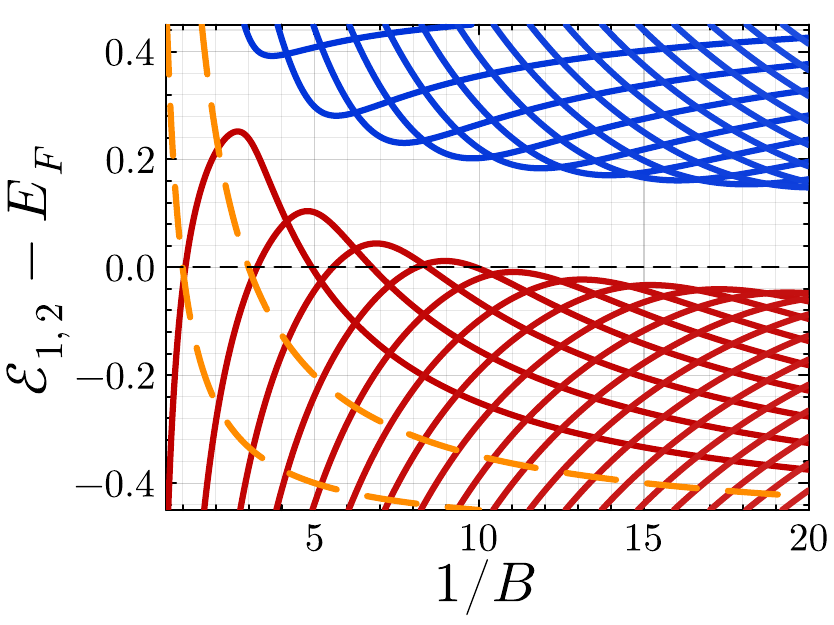}
        \put(0,70){(c)} % 在左上角添加标号
    \end{overpic}
    \caption{
        Landau-level spectrum $\mathcal{E}_{n}^{1,2}$. 
        (a,b) For $\delta\mu\leq0$, the LLs are monotonic in $1/B$, 
        and the system remains (almost) insulating. 
        (c) For $\delta\mu>0$, the LLs are non-monotonic in $1/B$, 
        and the system cross overs from metallic-like to insulating behavior as $1/B$ increases.
        The orange dashed curve shows the corresponding unhybridized LLs for comparison. 
        }
    \label{fig:LLstructures}
\end{figure}

For the two-band model, the Berry phase accumulated along a closed cyclotron 
orbit (see the supplemental material (SM)~\cite{SuppMat}) becomes $k$-dependent:
\begin{equation}
\Phi_{1,2}(k) =
\mp \pi l
\big[ 
1- \frac{\epsilon_d(k)-\epsilon_f(k)}
{\sqrt{(\epsilon_d(k)-\epsilon_f(k))^2+4V^2 k^{2l}}}
\big]. 
\label{eq:Phi_maintext_eps12}
\end{equation}
Since the phase of the off-diagonal hybridization term 
winds $l$ times around an isotropic orbit, 
the Berry phase on the inversion orbit for the limit of $\epsilon_d(k_F)=\epsilon_f(k_F)$, 
reduces to $\Phi_{1,2} = \mp l\pi$ for the two hybrid bands $\mathcal{E}_{1,2}(k)$, 
respectively. 
As the orbit moves away from the inversion point $k_F$, 
the Berry phase decreases in magnitude in the regime 
where the diagonal splitting dominates the hybridization, 
$|\epsilon_d(k)-\epsilon_f(k)|\gg 2|Vk^l|$ with $|k|\gg k_F$, 
as illustrated in Fig.~\ref{fig:BandInversion}(c).
This Berry phase distribution enters the phase shift 
of the quantum oscillations below.

%%%%%%

In the presence of a perpendicular magnetic field $B$, 
Landau quantization enters through the Peierls substitution, 
giving the dispersion for the Landau level (LL) index $n$,
\begin{align} 
\label{eq:LLeigen}
\begin{split}
\mathcal{E}_{n}^{1,2}
=&
\frac{1}{2}\big(\epsilon_{d,n}+\epsilon_{f,n-l}\big)
\\&
\pm
\frac{1}{2} \big[{(\epsilon_{d,n}-\epsilon_{f,n-l})^2
    +\frac{2^{l+2}n!}{(n-l)!} \frac{V^2 e^l B^l}{\hbar^l}}\big]^{1/2}
,
\end{split}
\end{align}
where ${n\geq l}$, ${\epsilon_{\lambda,n}=\hbar\omega_{\lambda}(n+1/2)-\mu_{\lambda}}$
for ${\lambda =d,f}$  is the unhybridized 
LL energy with cyclotron frequencies ${\omega_{\lambda}=eB/m_{\lambda}}$.
The HAM hybridization selectively couples only 
$\epsilon_{d,n}$ and $\epsilon_{f,n-l}$ with ${n\geq l}$, 
leaving the lowest $l$ LLs ($n=0,1,\ldots,l-1$) of each band decoupled.
The corresponding LL spectrum of the HAM band inversion is shown 
in Fig.~\ref{fig:LLstructures} for ${l=2}$, with more examples in SM~\cite{SuppMat}.
In particular, for the inverted insulator in Fig.~\ref{fig:LLstructures}(c), 
the LLs oscillate periodically in $1/B$, and the system crosses 
from a insulating regime at weak fields to a metallic-like regime 
at high field, separated by the critical field $B_c$. 
This is consistent with the understanding by comparing the cyclotron energy
and the hybridization band gap, and $B_c \approx \sqrt{m_d m_f}\Delta_D/e\hbar$~\cite{panda2022QuantumOscillationsMagnetization}. 
For ${B>B_c}$, the cyclotron energy overwhelms the hybridization band gap, and 
the LL spectra behave more like a metal in magnetic fields.

\begin{figure*}[t]
    \begin{overpic}[width=0.33\linewidth]{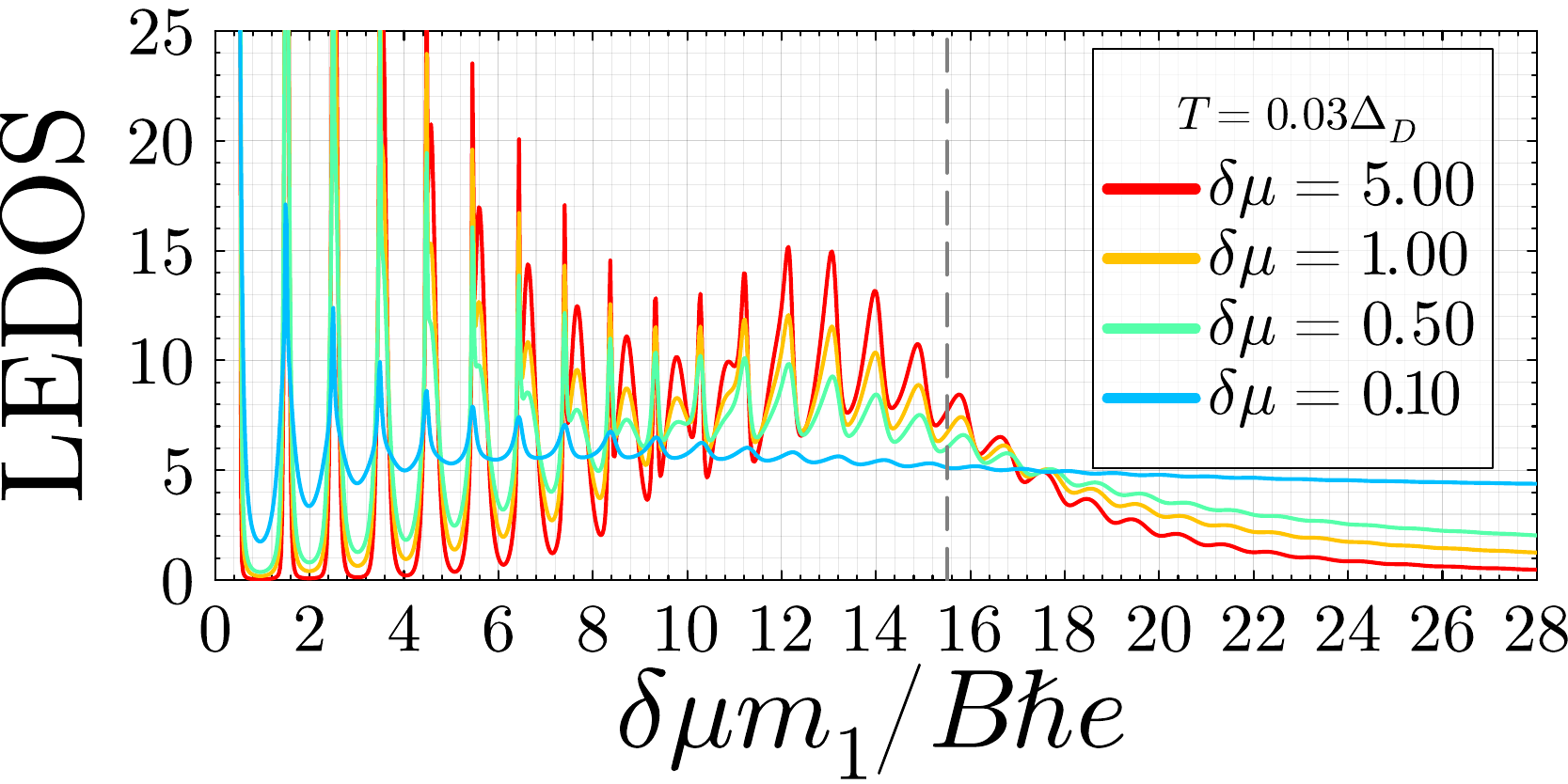}
        \put(0,45){(a)} % 在左上角添加标号
    \end{overpic}
    \begin{overpic}[width=0.33\linewidth]{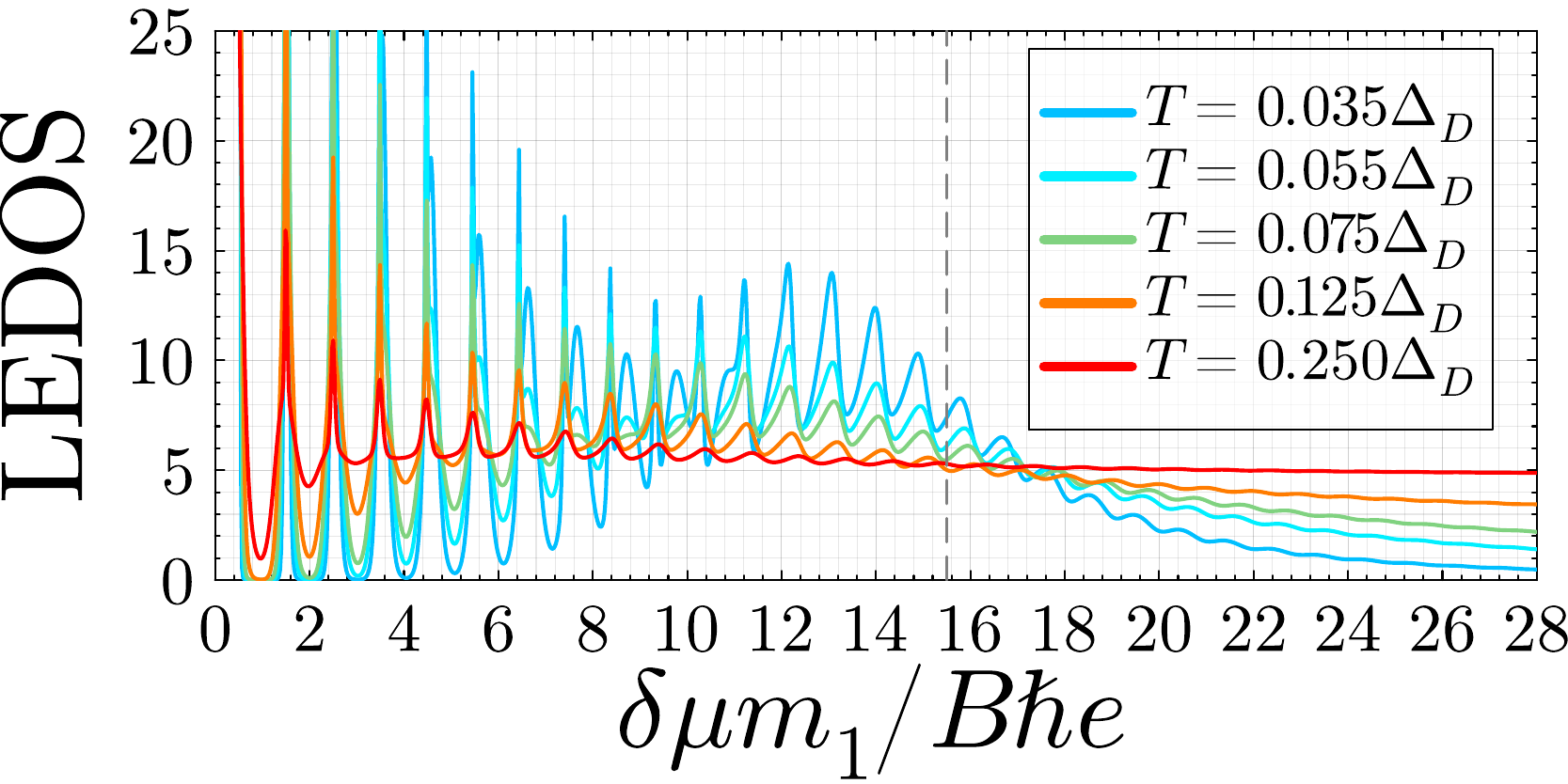}
        \put(0,45){(b)} % 在左上角添加标号
    \end{overpic}
    \begin{overpic}[width=0.32\linewidth]{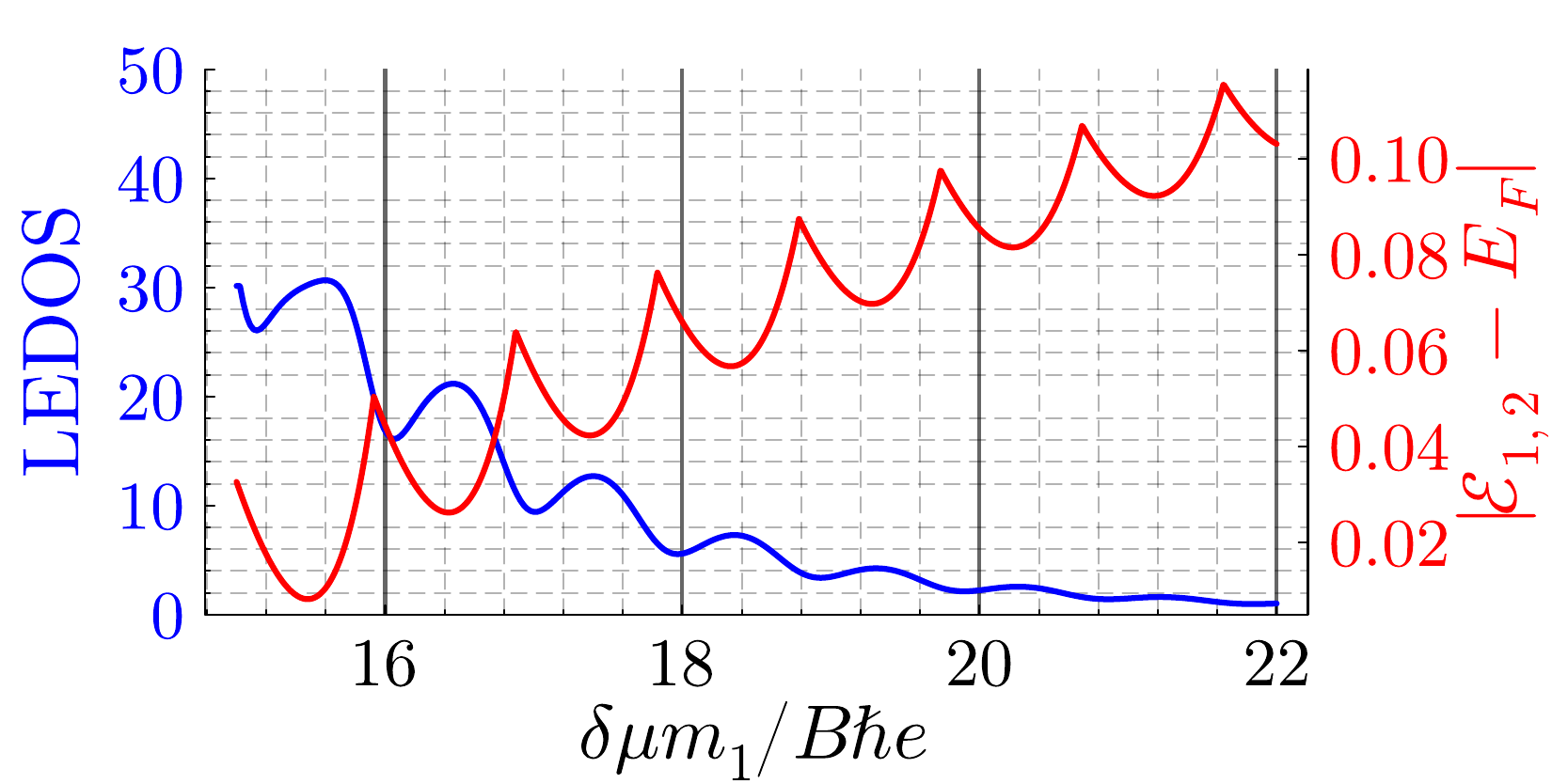}
        \put(0,45){(c)} % 在左上角添加标号
    \end{overpic}
    \caption{
        Quantum oscillations of the LEDOS for the two-band model with ${l=2}$.
        The oscillation behaviors for different detunings and temperatures 
        are shown in (a) and (b), respectively. The vertical dashed line marks 
        the critical field $B_c$. 
        (c) Comparison between the oscillation amplitude and $\min(|\mathcal{E}_{1,2}-E_F|)$ 
        at $T=0.035\Delta_D$. }
    \label{fig:LEDOS_Dmu_T_gap}
\end{figure*}

% 也可以横放一排

% \section{Low-Energy DOS}
% \label{sec:LEDOS}

\noindent\bluefour{\it Quantum oscillations.}---Here 
we analyze the quantum oscillations in the LEDOS near the chemical potential 
${\mu=E_F}$ for the two-band model. The LEDOS is given as 
\begin{equation} 
\label{eq:LEDOS}
    D(T)=\int_{-\infty}^{\infty}d\xi
    \frac{\partial n_\mathrm{F}(\xi-\mu,T)}{\partial \mu} A(\xi),
\end{equation}
where Fermi function $n_\mathrm{F}(\xi,T)=(e^{\beta\xi}+1)^{-1}$ 
with $\beta\!=\!1/T$  ($k_\mathrm{B}\!=\!1$),
and the single-particle DOS per unit area,
\begin{align} \label{eq:singleDOS}
    A(\xi) & =- {\rm Im}\sum_{n,b,p=\pm}\frac{ N_B / \pi}{\xi+E_F-\mathcal{E}_{n,bp} + i\Gamma},
\end{align}
with the LL degeneracy ${N_B=B e/2\pi\hbar}$ and $\Gamma\to0^+$. 

%%%

In the semimetals like Eq.~\eqref{eq:Model} with $V=0$ and $\delta\mu>0$, 
the electron and hole bands overlap and cross at $k_F$ and $E_F$,
and both bands contribute to quantum oscillations with the identical frequency 
${F_0 = (\hbar/2\pi e) \pi k_F^2}$, determined by the semiclassical orbit area at 
${E_F = \hbar^2 k_F^2 / 2m_d }$~\cite{zhang2016QuantumOscillationNarrowGap}. 
The LEDOS oscillations in this metallic limit arise from the periodic crossing of 
LL through the chemical potential ${\mu = E_F}$. 
Whenever ${\epsilon_n = \mu}$, the oscillation peak appears at the frequency $F_0$.
This follows the well-known Onsager's relation for quantum oscillations 
in metals~\cite{shoenberg1984MagneticOscillationsMetals}.

Using Eqs.~\eqref{eq:LEDOS} and \eqref{eq:singleDOS}, 
we examine the quantum oscillation of the LEDOS $D(T)$ 
for the hybridized LLs \eqref{eq:LLeigen} at different temperatures $T$ and $\delta\mu$. 
For the trivial band insulator with ${\delta\mu<0}$, all LLs bend away 
from the chemical potential ${\mu=E_F}$, yielding no crossing events 
in Fig.~\ref{fig:LLstructures}(a), and hence zero LEDOS. 
After the band inversion, the system for ${\delta\mu>0}$ 
enters the topological insulating regime of Eq.~\eqref{eq:Model}. As
we show in 
Fig.~\ref{fig:LEDOS_Dmu_T_gap}, clear quantum oscillations are  observed 
for different ${\delta\mu > 0}$ and temperatures $T$. 
In the insulating regime for small $B$'s ${B < B_c}$, the oscillations are  
tied to LLs approaching the band edges, which become thermally accessible at finite temperatures.
In Fig.~\ref{fig:LEDOS_Dmu_T_gap}, we further demonstrate that the oscillation amplitude tracks 
the minimum value of $|\mathcal{E}_{1,2}-E_F|$ at low temperatures, 
supporting this band-edge interpretation. In the metallic regime, however, 
the oscillation peaks are associated with LLs crossing the chemical potential $E_F$.

\begin{figure*}[t]
    \begin{minipage}[t]{0.25\textwidth}\vspace{0pt}
        \centering
        \begin{overpic}[width=\linewidth]{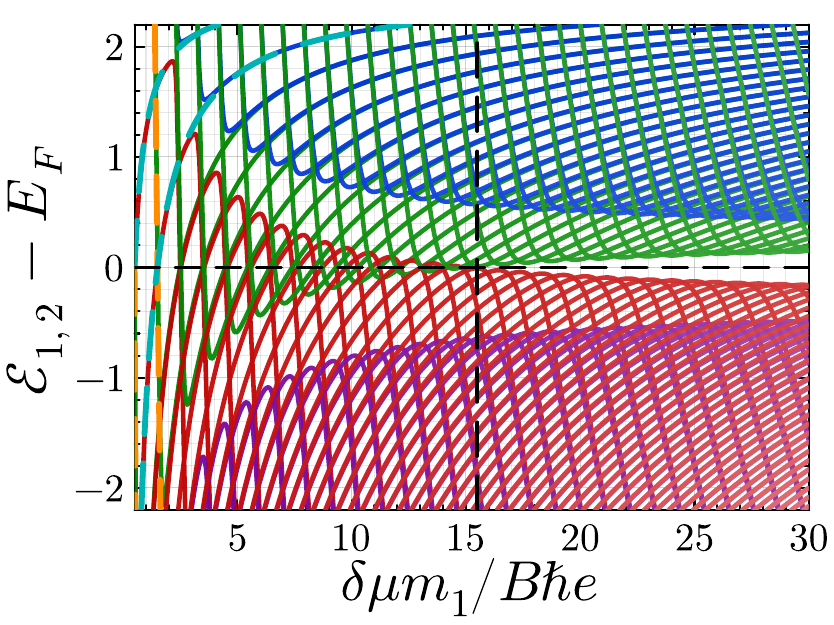}
            \put(1,70){(a)}
        \end{overpic}%
    \end{minipage}%\hfill
    \begin{minipage}[t]{0.25\textwidth}\vspace{0pt}
        \centering
        \begin{overpic}[width=\linewidth]{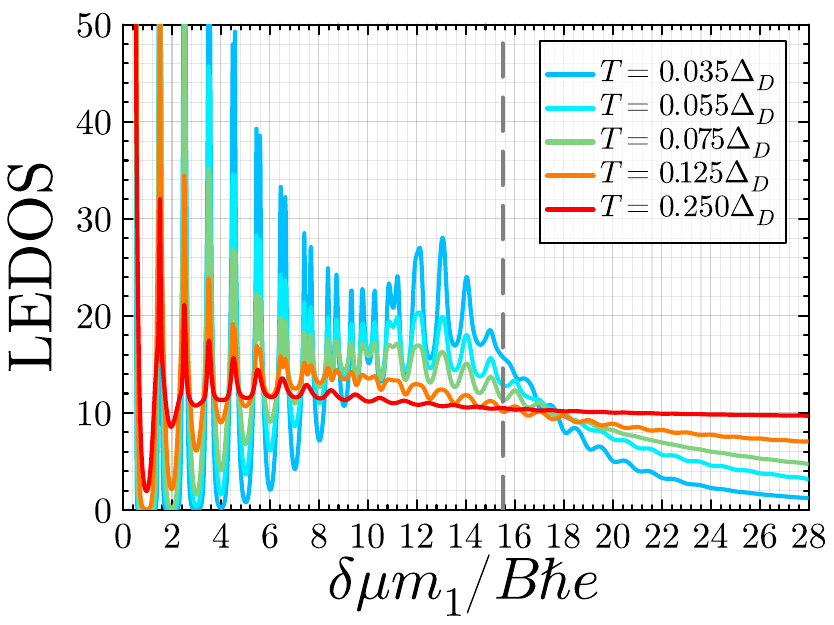}
            \put(-2,70){(b)}
        \end{overpic}
    \end{minipage}%\hfill
    \begin{minipage}[t][\dimexpr\ht0+\dp0\relax][t]{0.21\textwidth}\vspace{0pt}
        \centering
        \begin{overpic}[width=\linewidth,trim=0 90 0 0,clip]{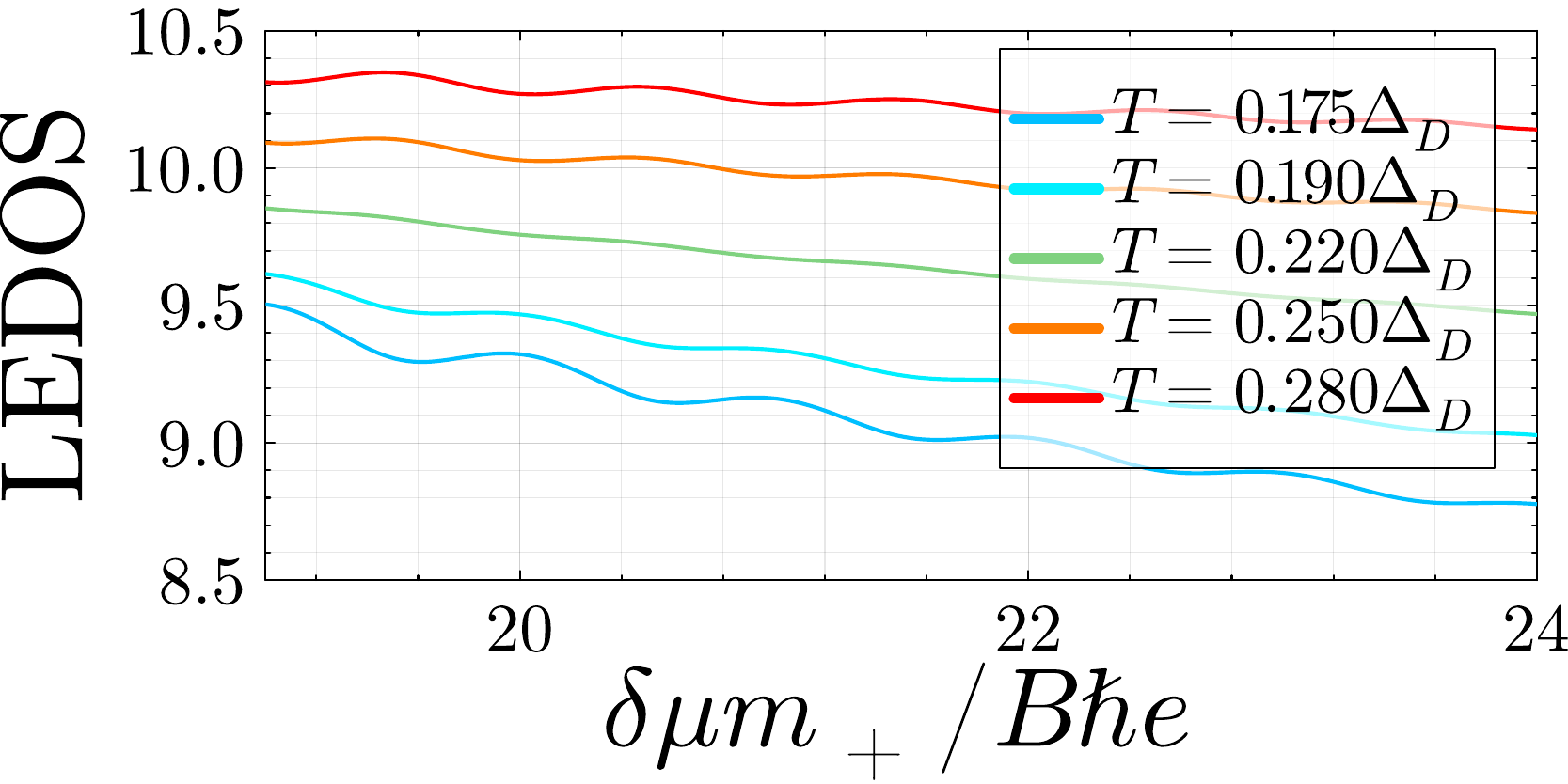}
            \put(-3,38){(c)}
        \end{overpic}
        % \vfill
        \begin{overpic}[width=\linewidth]{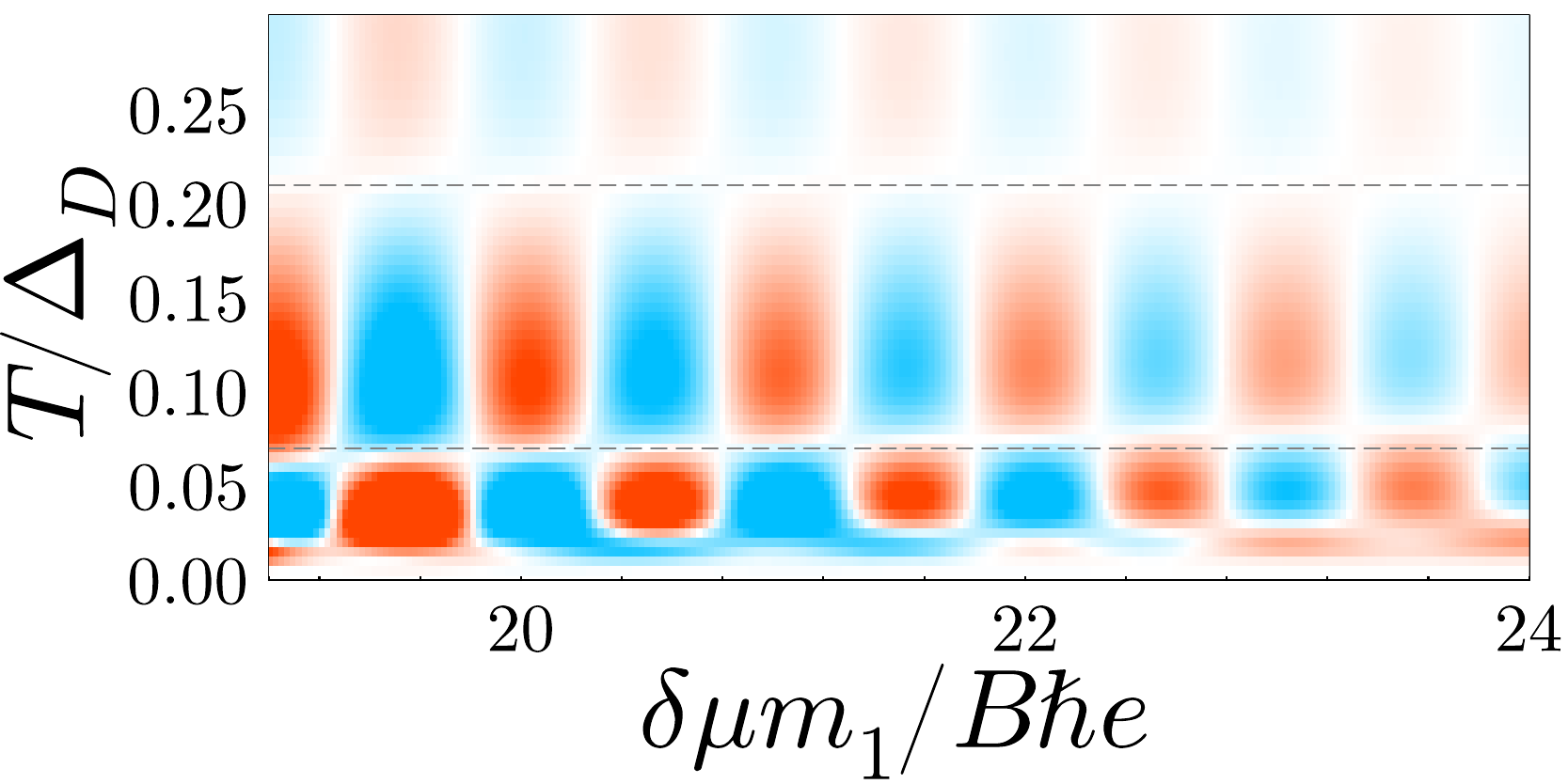}
            \put(-3,45){(d)}
        \end{overpic}
    \end{minipage}%\hfill
    \begin{minipage}[t]{0.25\textwidth}\vspace{0pt}
        \centering
        \begin{overpic}[width=\linewidth]{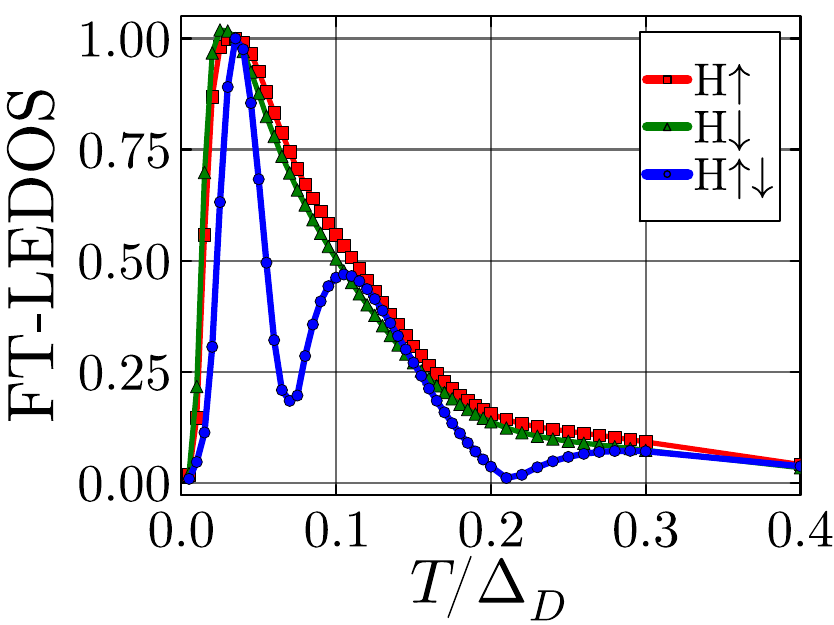}
            \put(0,70){(e)}
        \end{overpic}
    \end{minipage}
    \caption{
        (a) LL structures and
        (b) LEDOS oscillations of $H^{\uparrow\downarrow}_{\bm{k}}$.
        The vertical dashed line marks the critical field $B_c$. 
        (c) Thermal evolution of the oscillations near $T_c\approx 0.21\Delta_D$.
        (d) The corresponding detrended oscillations,
        showing $l=2$ characteristic temperatures. 
        (e) Temperature dependence of the normalized Fourier-transformed (FT) amplitude.
    }
    \label{fig:4band}
\end{figure*}

\noindent\bluefour{\it Four-band model.}---We now combine the two 
copies of the two-band models with the
opposite chiralities and convert them into
 a spinful Hamiltonian with the time-reversal symmetry.
This four-band model consists of one up-spin copy and one time-reversed down-spin copy,
and has a BHZ form as~\cite{bernevig2006QuantumSpin},  
\begin{equation}
    H^{\uparrow\downarrow}_{\bm{k}}
    = 
    \begin{pmatrix}
        H_{\uparrow} & 0 \\
        0 & H_{\downarrow}
    \end{pmatrix},
\end{equation}
where $H_{\uparrow} \equiv\mathcal{H}_{\bm{k}}$, 
$H_{\downarrow} \equiv\mathcal{H}_{-\bm{k}}^*$, 
and the basis vector becomes 
$\Psi_{\bm{k}} = (d_{\bm{k}\uparrow}, f_{\bm{k}\uparrow}, d_{\bm{k}\downarrow}, f_{\bm{k}\downarrow})^T$.
The time-reversal operator is $\mathcal{T} = i\sigma_y \mathcal{K}$ 
with $\sigma_y$ acting on the spin space and $\mathcal{K}$ the complex conjugation.
Each spin block realizes a Chern-insulator-like band inversion with the opposite Chern numbers, 
while the full time-reversal-invariant four-band model describes a transition from a normal insulator to a quantum spin Hall insulator characterized by helical edge modes.

Since the off-block-diagonal part of $H^{\uparrow\downarrow}_{\bm{k}}$ is zero, 
the LL spectrum of the four-band model is a simple superposition of 
the time-reversal-related blocks $H_{\uparrow\downarrow}$, as 
it is shown in Fig.~\ref{fig:4band}(a). 
The total LEDOS is thus the sum of their oscillatory contributions 
[see Fig.~\ref{fig:4band}(b)].
Despite the simple summation,
the two blocks actually acquire the opposite temperature-dependent phase shifts 
in the insulating regime ${B<B_c}$~\cite{SuppMat}, 
producing the interference pattern in Fig.~\ref{fig:4band}(c,d). 
The total signal takes the form~\cite{SuppMat}, 
\begin{equation}
    D_{\mathrm{tot}}^{\mathrm{osc}}(B,T)
    \simeq
    4A(T)
    \cos\! \Big[\frac{\Delta\phi(T)}{2}\Big]
    \cos\!\Big[2\pi\frac{F_0}{B}+\bar\phi(T)\Big],
    \label{eq:main_interference_phase}
\end{equation}
where 
${\Delta\phi(T)=\phi_{\uparrow}(T)-\phi_{\downarrow}(T)}$  
and  
$\bar\phi(T)=\frac{1}{2}[\phi_{\uparrow}(T)+\phi_{\downarrow}(T)]$. 
Here, $\phi_{\uparrow}(T)$ and $\phi_{\downarrow}(T)$ are not the Berry phases of single cyclotron orbits. 
Instead, they are the phases of the thermally weighted complex amplitudes of the two time-reversal-related blocks~\cite{SuppMat}, into which the energy-resolved Berry phases $\Phi_{\lambda,\uparrow}(E)$ and $\Phi_{\lambda,\downarrow}(E)$ enter through the band-edge orbits. 
Because time reversal requires $\Phi_{\lambda,\uparrow}(E)=-\Phi_{\lambda,\downarrow}(E)$, the two effective phases evolve in opposite directions as the thermal window broadens away from the inversion orbit. 
Consequently, $\Delta\phi(T)$ measures the phase mismatch between the two blocks and controls the interference envelope through $\cos[\Delta\phi(T)/2]$, whereas $\bar\phi(T)$ sets the phase offset of the resulting oscillation. 
In the symmetric limit $\phi_{\uparrow}(T)=-\phi_{\downarrow}(T)$, one has $\bar\phi(T)=0$, so the suppression is governed primarily by the temperature dependence of $\Delta\phi(T)$.
Destructive interference occurs near the characteristic temperatures 
$T_c$ at which ${\Delta\phi(T_c)\approx(2j+1)\pi}$, thus the leading harmonics 
are nearly cancelled out. 
This explains the strong suppression and the phase-slip structure that is found 
in Fig.~\ref{fig:4band}(c,d).  
In the present ${l=2}$ case, 
the number of the characteristic temperatures matches with $l$
and is consistent with the semiclassical picture that 
the unwrapped relative phase $\Delta\phi(T)$ sweeps an $l$-dependent range as 
the temperature increases.

The same interpretation is further supported by the Fourier spectra of the LEDOS shown in Fig.~\ref{fig:4band}(e). 
Destructive interference between the two time-reversal-related blocks drives the spectral weight of the total LEDOS to zero at the characteristic temperatures. 
Taking the direct gap $\Delta_D$ as the temperature scale, the suppression occurs at approximately $0.21\Delta_D$ and $0.07\Delta_D$ for ${l=2}$, and at $0.21\Delta_D$, $0.04\Delta_D$, and $0.09\Delta_D$ for ${l=3}$~\cite{SuppMat}.
This non-monotonic temperature dependence of the oscillation amplitude, 
together with the associated phase evolution, 
provides a direct experimental signature of HAM band inversion 
and clearly distinguishes it from the monotonic Lifshitz-Kosevich behavior expected 
for a simple metal~\cite{zhang2016QuantumOscillationNarrowGap,panda2022QuantumOscillationsMagnetization}.

% \section{Discussion and conclusion}

% 

\noindent\bluefour{\it Discussion.}---For the single chiral block 
of a two band model with the HAM, the 
Landau quantization in the inverted insulating regime is sufficient 
to generate the thermally activated quantum oscillations in LEDOS, 
with a phase that acquires a non-trivial temperature dependence 
through the orbit-dependent Berry phase. When the two opposite 
chiralities are combined into a time-reversal-invariant four-band model, 
their interference produces a clear departure from 
the conventional Lifshitz-Kosevich behavior, manifested 
by pronounced suppression of the leading oscillation amplitude 
at the characteristic temperatures.

These results provide the experimentally accessible signatures 
of both the HAM hybridization and the underlying inverted insulators 
in magnetic fields. 
They are especially relevant to the moir\'e and other engineered platforms, 
where the hybridization gap and band inversion can be tuned by 
the gate voltage or strain, and where the reduced $B_c$ at larger $l$'s 
provides a practical route between activated 
and metallic-like oscillation regimes~\cite{li2021QuantumAnomalous,zhang2021SpintexturedChern}. 
Finally, our analysis could have implications for future studies 
of the correlation-driven effects in the band-inverted systems, 
including various excitonic instabilities, 
their consequences for the band edges 
and the quantum-oscillation spectrum,
and even the fractionalized excitonic effects~\cite{nguyen2026QuantumOscillations,qi2026CompetitionExcitonic,hu2018FractionalExcitonicInsulator}.
The non-trivial combination and interference between the 
time-reversal-related HAM band inversion blocks in our work can be naturally extended 
to other discrete-symmetry-related HAM electron-hole hybridizations and excitonic 
pairings with and without spin-orbit coupling, 
which may provide a useful framework for interpreting complicated oscillatory responses in related materials.

% \section*{Acknowledgements}

\noindent\bluefour{\it Acknowledgments.}---J.Y. thanks Lingxian Kong 
and Junyu Tang for helpful discussions.
This work is supported by NSFC with Grants No.~92565110 and No.~12574061, 
BJNSF with No.~F261004 and from Quantum Science and Technology-National 
Science and Technology Major Project (grant No.~2025ZD0300500). 

\bibstyle{apsrev-nourl}
\bibliography{Refs}

@misc{SuppMat,
title = {{See Supplemental Material for additional supporting results.}}
}

@article{zhang2016QuantumOscillationNarrowGap,
  title = {Quantum {{Oscillation}} in {{Narrow-Gap Topological Insulators}}},
  author = {Zhang, Long and Song, Xue-Yang and Wang, Fa},
  year = 2016,
  journal = {Phys. Rev. Lett.},
  volume = {116},
  number = {4},
  pages = {046404},
  issn = {0031-9007, 1079-7114},
  doi = {10.1103/PhysRevLett.116.046404},
  copyright = {http://link.aps.org/licenses/aps-default-license}
}

@article{xu2012SpinLiquid,
  title = {Spin {{Liquid Phases}} for {{Spin-1 Systems}} on the {{Triangular Lattice}}},
  author = {Xu, Cenke and Wang, Fa and Qi, Yang and Balents, Leon and Fisher, Matthew P. A.},
  year = 2012,
  month = feb,
  journal = {Phys. Rev. Lett.},
  volume = {108},
  number = {8},
  pages = {087204},
  publisher = {American Physical Society},
  doi = {10.1103/PhysRevLett.108.087204}
}

@article{li2021QuantumAnomalous,
  title = {Quantum Anomalous {{Hall}} Effect from Intertwined Moir\'e Bands},
  author = {Li, Tingxin and Jiang, Shengwei and Shen, Bowen and Zhang, Yang and Li, Lizhong and Tao, Zui and Devakul, Trithep and Watanabe, Kenji and Taniguchi, Takashi and Fu, Liang and Shan, Jie and Mak, Kin Fai},
  year = 2021,
  month = dec,
  journal = {Nature},
  volume = {600},
  number = {7890},
  pages = {641--646},
  publisher = {Nature Publishing Group},
  issn = {1476-4687},
  doi = {10.1038/s41586-021-04171-1},
  copyright = {2021 The Author(s), under exclusive licence to Springer Nature Limited},
  langid = {english}
}

@article{zhang2021SpintexturedChern,
  title = {Spin-Textured {{Chern}} Bands in {{AB-stacked}} Transition Metal Dichalcogenide Bilayers},
  author = {Zhang, Yang and Devakul, Trithep and Fu, Liang},
  year = 2021,
  month = sep,
  journal = {Proc. Natl. Acad. Sci. U.S.A.},
  volume = {118},
  number = {36},
  pages = {e2112673118},
  publisher = {Proceedings of the National Academy of Sciences},
  doi = {10.1073/pnas.2112673118},
  langid = {english}
}

@article{mikitik2004BerryPhase,
  title = {{Berry Phase and de Haas--van Alphen Effect in ${\mathrm{L}\mathrm{a}\mathrm{R}\mathrm{h}\mathrm{I}\mathrm{n}}_{5}$}},
  author = {Mikitik, G. P. and Sharlai, {\relax Yu}. V.},
  year = 2004,
  month = sep,
  journal = {Phys. Rev. Lett.},
  volume = {93},
  number = {10},
  pages = {106403},
  issn = {0031-9007, 1079-7114},
  doi = {10.1103/PhysRevLett.93.106403},
  copyright = {http://link.aps.org/licenses/aps-default-license},
  langid = {english}
}

@article{zhang2005ExperimentalObservation,
  title = {Experimental Observation of the Quantum {{Hall}} Effect and {{Berry}}'s Phase in Graphene},
  author = {Zhang, Yuanbo and Tan, Yan-Wen and Stormer, Horst L. and Kim, Philip},
  year = 2005,
  month = nov,
  journal = {Nature},
  volume = {438},
  number = {7065},
  pages = {201--204},
  publisher = {Nature Publishing Group},
  issn = {1476-4687},
  doi = {10.1038/nature04235},
  copyright = {2005 Springer Nature Limited},
  langid = {english}
}

@book{shoenberg1984MagneticOscillationsMetals,
  title = {Magnetic {{Oscillations}} in {{Metals}}},
  author = {Shoenberg, D.},
  year = {1984},
  series = {Cambridge {{Monographs}} on {{Physics}}},
  publisher = {Cambridge University Press},
  address = {Cambridge},
  doi = {10.1017/CBO9780511897870},
  isbn = {978-0-521-11878-1}
}

@article{xiang2018QuantumOscillations,
  title = {Quantum Oscillations of Electrical Resistivity in an Insulator},
  author = {Xiang, Z. and Kasahara, Y. and Asaba, T. and Lawson, B. and Tinsman, C. and Chen, Lu and Sugimoto, K. and Kawaguchi, S. and Sato, Y. and Li, G. and Yao, S. and Chen, Y. L. and Iga, F. and Singleton, John and Matsuda, Y. and Li, Lu},
  year = 2018,
  month = oct,
  journal = {Science},
  volume = {362},
  number = {6410},
  pages = {65--69},
  publisher = {American Association for the Advancement of Science},
  doi = {10.1126/science.aap9607}
}

@misc{navarro-labastida2025TopologicalPhasea,
  title = {Topological Phase Diagram of Twisted Bilayer Graphene as a Function of the Twist Angle},
  author = {{Navarro-Labastida}, Leonardo A. and Pantaleon, Pierre A. and Guinea, Francisco and Naumis, Gerardo G.},
  year = 2025,
  month = jul,
  number = {arXiv:2507.05341},
  eprint = {2507.05341},
  doi = {10.48550/arXiv.2507.05341},
  publisher = {arXiv},
  archiveprefix = {arXiv},
  langid = {english}
}

@article{allocca2024FluctuationdominatedQuantumOscillations,
  title = {Fluctuation-Dominated Quantum Oscillations in Excitonic Insulators},
  author = {Allocca, Andrew A. and Cooper, Nigel R.},
  year = 2024,
  month = {Aug},
  journal = {Phys. Rev. Res.},
  volume = {6},
  issue = {3},
  numpages = {11},
  pages = {033199},
  doi = {10.1103/PhysRevResearch.6.033199},
}

@article{motrunich2006OrbitalMagnetic,
  title = {{Orbital magnetic field effects in spin liquid with spinon Fermi sea: Possible application to $\ensuremath{\kappa}\text{\ensuremath{-}}{(\mathrm{ET})}_{2}{\mathrm{Cu}}_{2}{(\mathrm{C}\mathrm{N})}_{3}$}},
  shorttitle = {Orbital Magnetic Field Effects in Spin Liquid with Spinon {{Fermi}} Sea},
  author = {Motrunich, Olexei I.},
  year = 2006,
  month = apr,
  journal = {Phys. Rev. B},
  volume = {73},
  number = {15},
  pages = {155115},
  publisher = {American Physical Society},
  doi = {10.1103/PhysRevB.73.155115}
}

@article{chowdhury2018MixedvalenceInsulators,
  title = {{Mixed-Valence Insulators with Neutral {{Fermi}} Surfaces}},
  author = {Chowdhury, Debanjan and Sodemann, Inti and Senthil, T.},
  year = 2018,
  month = may,
  journal = {Nat Commun},
  volume = {9},
  number = {1},
  pages = {1766},
  publisher = {Nature Publishing Group},
  issn = {2041-1723},
  doi = {10.1038/s41467-018-04163-2},
  copyright = {2018 The Author(s)},
  langid = {english}
}

@article{sodemann2018QuantumOscillationsa,
  title = {{Quantum Oscillations in Insulators with Neutral {{Fermi}} Surfaces}},
  author = {Sodemann, Inti and Chowdhury, Debanjan and Senthil, T.},
  year = 2018,
  month = jan,
  journal = {Phys. Rev. B},
  volume = {97},
  number = {4},
  pages = {045152},
  publisher = {American Physical Society},
  doi = {10.1103/PhysRevB.97.045152}
}

@article{davenport2026BerryCurvature,
  title = {{Berry Curvature of Low-Energy Excitons in Rhombohedral Graphene}},
  author = {Davenport, Henry},
  year = 2026,
  journal = {Phys. Rev. B},
  volume = {113},
  number = {11},
  doi = {10.1103/shrm-4swg}
}

@article{dong2024StabilityAnomalous,
  title = {{Stability of Anomalous {{Hall}} Crystals in Multilayer Rhombohedral Graphene}},
  author = {Dong, Zhihuan and Patri, Adarsh S. and Senthil, T.},
  year = 2024,
  month = nov,
  journal = {Phys. Rev. B},
  volume = {110},
  number = {20},
  pages = {205130},
  publisher = {American Physical Society},
  doi = {10.1103/PhysRevB.110.205130}
}

@misc{thiagarajan2025NatureTopological,
  title = {Nature of the {{Topological Transition}} of the {{Kitaev Model}} in [111] {{Magnetic Field}}},
  author = {Thiagarajan, S. and Watson, C. and Yzeiri, T. and Hu, H. and Uchoa, B. and Kr{\"u}ger, F.},
  year = 2025,
  month = sep,
  number = {arXiv:2509.13057},
  eprint = {2509.13057},
  primaryclass = {cond-mat},
  publisher = {arXiv},
  doi = {10.48550/arXiv.2509.13057},
  archiveprefix = {arXiv}
}

@article{zhang2022TheoryKitaevModel,
  title = {Theory of the {{Kitaev}} Model in a [111] Magnetic Field},
  author = {Zhang, Shang-Shun and Hal{\'a}sz, G{\'a}bor B. and Batista, Cristian D.},
  year = 2022,
  month = jan,
  journal = {Nat Commun},
  volume = {13},
  number = {1},
  pages = {399},
  publisher = {Nature Publishing Group},
  issn = {2041-1723},
  doi = {10.1038/s41467-022-28014-3},
  copyright = {2022 The Author(s)},
  langid = {english}
}

@article{zhang2015SpontaneousChiral,
  title = {Spontaneous Chiral Symmetry Breaking in Bilayer Graphene},
  author = {Zhang, Fan},
  year = 2015,
  journal = {Synthetic Metals},
  volume = {210},
  pages = {9--18},
  issn = {0379-6779},
  doi = {10.1016/j.synthmet.2015.07.028}
}

@article{gassner2025FluxAttachmentTheory,
  title = {Flux {{Attachment Theory}} of {{Fractional Excitonic Insulators}}},
  author = {Gassner, Steven and Stern, Ady and Kane, C. L.},
  year = 2025,
  month = oct,
  journal = {Phys. Rev. Lett.},
  volume = {135},
  number = {18},
  pages = {186602},
  publisher = {American Physical Society},
  doi = {10.1103/x789-kxy3}
}

@article{bernevig2006QuantumSpin,
  title = {Quantum {{Spin Hall Effect}} and {{Topological Phase Transition}} in {{HgTe Quantum Wells}}},
  author = {Bernevig, B. Andrei and Hughes, Taylor L. and Zhang, Shou-Cheng},
  year = 2006,
  month = dec,
  journal = {Science},
  volume = {314},
  number = {5806},
  pages = {1757--1761},
  publisher = {American Association for the Advancement of Science},
  doi = {10.1126/science.1133734}
}

@article{qi2026CompetitionExcitonic,
  title = {Competition between Excitonic Insulators and Quantum {{Hall}} States in Correlated Electron--Hole Bilayers},
  author = {Qi, Ruishi and Li, Qize and Zhang, Zuocheng and Nie, Jiahui and Zou, Bo and Cui, Zhiyuan and Kim, Haleem and Sanborn, Collin and Chen, Sudi and Xie, Jingxu and Taniguchi, Takashi and Watanabe, Kenji and Crommie, Michael F. and MacDonald, Allan H. and Wang, Feng},
  year = 2026,
  month = jan,
  journal = {Nat. Mater.},
  volume = {25},
  number = {1},
  pages = {35--41},
  publisher = {Nature Publishing Group},
  issn = {1476-4660},
  doi = {10.1038/s41563-025-02316-5},
  copyright = {2025 The Author(s), under exclusive licence to Springer Nature Limited},
  langid = {english}
}

@article{nguyen2026QuantumOscillations,
  title = {Quantum Oscillations in a Dipolar Excitonic Insulator},
  author = {Nguyen, Phuong X. and Chaturvedi, Raghav and Zou, Bo and Watanabe, Kenji and Taniguchi, Takashi and MacDonald, Allan H. and Mak, Kin Fai and Shan, Jie},
  year = 2026,
  month = jan,
  journal = {Nat. Mater.},
  volume = {25},
  number = {1},
  pages = {42--48},
  publisher = {Nature Publishing Group},
  issn = {1476-4660},
  doi = {10.1038/s41563-025-02334-3},
  copyright = {2025 The Author(s)},
  langid = {english}
}

@article{knolle2015QuantumOscillationsFermi,
  title = {Quantum {{Oscillations}} without a {{Fermi Surface}} and the {{Anomalous}} de {{Haas}}--van {{Alphen Effect}}},
  author = {Knolle, Johannes and Cooper, Nigel R.},
  year = 2015,
  month = sep,
  journal = {Phys. Rev. Lett.},
  volume = {115},
  number = {14},
  pages = {146401},
  issn = {0031-9007, 1079-7114},
  doi = {10.1103/PhysRevLett.115.146401},
  copyright = {http://link.aps.org/licenses/aps-default-license},
  langid = {english}
}

@article{moon2013NonFermiLiquidTopological,
  title = {Non-{{Fermi-Liquid}} and {{Topological States}} with {{Strong Spin-Orbit Coupling}}},
  author = {Moon, Eun-Gook and Xu, Cenke and Kim, Yong Baek and Balents, Leon},
  year = 2013,
  month = nov,
  journal = {Phys. Rev. Lett.},
  volume = {111},
  number = {20},
  pages = {206401},
  issn = {0031-9007, 1079-7114},
  doi = {10.1103/PhysRevLett.111.206401},
  copyright = {http://link.aps.org/licenses/aps-default-license},
  langid = {english}
}

@article{yao2018LuttingerSemimetal,
  title = {${\mathrm{Pr}}_{2}{\mathrm{Ir}}_{2}{\mathrm{O}}_{7}$: When Luttinger Semimetal Meets Melko-Hertog-Gingras Spin Ice State},
  author = {Yao, Xu-Ping and Chen, Gang},
  year = 2018,
  month = dec,
  journal = {Phys. Rev. X},
  volume = {8},
  number = {4},
  pages = {041039},
  publisher = {American Physical Society},
  doi = {10.1103/PhysRevX.8.041039}
}

@article{knolle2017AnomalousHaas,
  title = {Anomalous de {{Haas}}--van {{Alphen Effect}} in {{InAs}} / {{GaSb Quantum Wells}}},
  author = {Knolle, Johannes and Cooper, Nigel R.},
  year = {2017},
  month = apr,
  journal = {Phys. Rev. Lett.},
  volume = {118},
  number = {17},
  pages = {176801},
  issn = {0031-9007, 1079-7114},
  doi = {10.1103/PhysRevLett.118.176801},
  copyright = {http://link.aps.org/licenses/aps-default-license},
  langid = {english}
}

@article{venderbos2018HigherAngularMomentum,
  title = {Higher Angular Momentum Band Inversions in Two Dimensions},
  author = {Venderbos, J{\"o}rn W. F. and Hu, Yichen and Kane, C. L.},
  year = 2018,
  journal = {Phys. Rev. B},
  volume = {98},
  number = {23},
  pages = {235160},
  publisher = {American Physical Society},
  doi = {10.1103/PhysRevB.98.235160}
}

@article{hu2018FractionalExcitonicInsulator,
  title = {Fractional {{Excitonic Insulator}}},
  author = {Hu, Yichen and Venderbos, J{\"o}rn W. F. and Kane, C. L.},
  year = 2018,
  journal = {Phys. Rev. Lett.},
  volume = {121},
  number = {12},
  pages = {126601},
  issn = {0031-9007, 1079-7114},
  doi = {10.1103/PhysRevLett.121.126601}
}

@article{panda2022QuantumOscillationsMagnetization,
  title = {Quantum Oscillations in the Magnetization and Density of States of Insulators},
  author = {Panda, Animesh and Banerjee, Sumilan and Randeria, Mohit},
  year = 2022,
  journal = {Proc. Natl. Acad. Sci. U.S.A.},
  volume = {119},
  number = {42},
  pages = {e2208373119},
  publisher = {Proceedings of the National Academy of Sciences},
  doi = {10.1073/pnas.2208373119},
  copyright = {Copyright {\copyright} 2022 the Author(s). Published by PNAS.}
}

\clearpage
\newpage
% \appendix
\widetext

\begin{center}
\textbf{\large Supplemental Material for \\``Quantum oscillations in proximity to high-angular-momentum band inversion''}
\end{center}

% \tableofcontents
\addtocontents{toc}{\protect\setcounter{tocdepth}{0}}
{
\tableofcontents
}

\renewcommand{\thefigure}{S\arabic{figure}}
\setcounter{figure}{0}
\renewcommand{\theequation}{S\arabic{equation}}
\setcounter{equation}{0}
\renewcommand{\thesection}{\Roman{section}}
\setcounter{section}{0}
\setcounter{secnumdepth}{4}

\section{Berry phase of a generic two-band model}
\label{app:berry_phase_eps1_eps2}

In this Appendix, we derive the Berry phase for a generic two-band Hamiltonian with unequal diagonal dispersions. 
This provides a convenient unified form for discussing Dirac fermion, and band-inverted models with high angular momentum.

We consider a two-band Hamiltonian of the form
\begin{equation}
H(\mathbf{k})=
\begin{pmatrix}
\epsilon_1(\mathbf{k}) & g(\mathbf{k})\\
g^*(\mathbf{k}) & \epsilon_2(\mathbf{k})
\end{pmatrix},
\label{eq:H_eps12}
\end{equation}
where $\epsilon_1(\mathbf{k})$ and $\epsilon_2(\mathbf{k})$ are the diagonal dispersions, and $g(\mathbf{k})$ is the generally complex hybridization matrix element. It is convenient to introduce
\begin{equation}
\bar{\epsilon}(\mathbf{k})=\frac{\epsilon_1(\mathbf{k})+\epsilon_2(\mathbf{k})}{2},
\qquad
\Delta(\mathbf{k})=\frac{\epsilon_1(\mathbf{k})-\epsilon_2(\mathbf{k})}{2},
\label{eq:barDelta}
\end{equation}
so that the Hamiltonian becomes
\begin{equation}
H(\mathbf{k})
=
\bar{\epsilon}(\mathbf{k})\,\mathbbm{1}
+
\begin{pmatrix}
\Delta(\mathbf{k}) & g(\mathbf{k})\\
g^*(\mathbf{k}) & -\Delta(\mathbf{k})
\end{pmatrix}.
\label{eq:H_split}
\end{equation}
Writing the off-diagonal term as
\begin{equation}
g(\mathbf{k})=\rho(\mathbf{k})e^{-i\phi(\mathbf{k})},
\qquad
\rho(\mathbf{k})=|g(\mathbf{k})|,
\label{eq:gpolar}
\end{equation}
we may cast the Hamiltonian into the standard pseudospin form
\begin{equation}
H(\mathbf{k})=
\bar{\epsilon}(\mathbf{k})\,\mathbbm{1}
+
\mathbf{d}(\mathbf{k})\cdot\boldsymbol{\sigma},
\label{eq:H_dvec}
\end{equation}
with
$\mathbf{d}(\mathbf{k})=
\bigl(
\rho(\mathbf{k})\cos\phi(\mathbf{k}),
\,
\rho(\mathbf{k})\sin\phi(\mathbf{k}),
\,
\Delta(\mathbf{k})
\bigr)$.
The eigenvalues are therefore
\begin{equation}
E_\pm(\mathbf{k})=
\bar{\epsilon}(\mathbf{k})\pm \varepsilon(\mathbf{k}),
\qquad
\varepsilon(\mathbf{k})=
\sqrt{\Delta(\mathbf{k})^2+\rho(\mathbf{k})^2}.
\label{eq:eigs_eps}
\end{equation}
Since the Berry phase depends only on the eigenvectors, the scalar term $\bar{\epsilon}(\mathbf{k})\mathbbm{1}$ plays no role in the derivation below. Thus, the Berry phase is insensitive to the average diagonal energy and depends only on the normalized pseudospin texture $\hat{\mathbf d}=\mathbf d/|\mathbf d|$.

To parameterize the pseudospin direction, we introduce an angle $\chi(\mathbf{k})$ through
\begin{equation}
\cos\chi(\mathbf{k})=\frac{\Delta(\mathbf{k})}{\varepsilon(\mathbf{k})},
\qquad
\sin\chi(\mathbf{k})=\frac{\rho(\mathbf{k})}{\varepsilon(\mathbf{k})},
\label{eq:chi_eps}
\end{equation}
such that the normalized pseudospin vector can be expressed as
$\hat{\mathbf d}(\mathbf{k})
=
\bigl(
\sin\chi \cos\phi,\,
\sin\chi \sin\phi,\,
\cos\chi
\bigr)$.
A convenient gauge choice for the normalized eigenvectors is
\begin{equation}
|u_+(\mathbf{k})\rangle=
\begin{pmatrix}
\cos\frac{\chi}{2}\\[4pt]
e^{i\phi}\sin\frac{\chi}{2}
\end{pmatrix},
\qquad
|u_-(\mathbf{k})\rangle=
\begin{pmatrix}
-\,e^{-i\phi}\sin\frac{\chi}{2}\\[4pt]
\cos\frac{\chi}{2}
\end{pmatrix},
\label{eq:evecs_eps}
\end{equation}
corresponding to the upper and lower bands $E_+$ and $E_-$, respectively.

The Berry connection is defined by
\begin{equation}
\mathcal{A}_\pm(\mathbf{k})
=
i\langle u_\pm(\mathbf{k})|\nabla_{\mathbf{k}}u_\pm(\mathbf{k})\rangle.
\label{eq:Apm_def}
\end{equation}
Using Eq.~\eqref{eq:evecs_eps}, one finds
\begin{equation}
\mathcal{A}_+(\mathbf{k})
=
-\frac{1-\cos\chi(\mathbf{k})}{2}\,\nabla_{\mathbf{k}}\phi(\mathbf{k}),
\quad
\mathcal{A}_-(\mathbf{k})
=
+\frac{1-\cos\chi(\mathbf{k})}{2}\,\nabla_{\mathbf{k}}\phi(\mathbf{k}),
\label{eq:Aplus_eps_Aminus_eps}
\end{equation}
up to the same gauge convention.

Substituting Eq.~\eqref{eq:chi_eps}, these expressions become
\begin{equation}
\mathcal{A}_+(\mathbf{k})
=
-\frac{1}{2}
\left[
1-\frac{\Delta(\mathbf{k})}{\sqrt{\Delta(\mathbf{k})^2+\rho(\mathbf{k})^2}}
\right]
\nabla_{\mathbf{k}}\phi(\mathbf{k}),
\quad
\mathcal{A}_-(\mathbf{k})
=
+\frac{1}{2}
\left[
1-\frac{\Delta(\mathbf{k})}{\sqrt{\Delta(\mathbf{k})^2+\rho(\mathbf{k})^2}}
\right]
\nabla_{\mathbf{k}}\phi(\mathbf{k}).
\label{eq:Aplus_explicit_Aminus_explicit}
\end{equation}

The Berry phase accumulated along a closed orbit $C$ in momentum space is
\begin{equation}
\Phi_\pm[C]
=
\oint_C \mathcal{A}_\pm(\mathbf{k})\cdot d\mathbf{k}.
\label{eq:Phi_def_closed}
\end{equation}
Using Eqs.~\eqref{eq:Aplus_explicit_Aminus_explicit}, we obtain
\begin{equation}
\Phi_\pm[C]
=
\mp \frac{1}{2}
\oint_C
\left[
1-\frac{\Delta(\mathbf{k})}{\sqrt{\Delta(\mathbf{k})^2+\rho(\mathbf{k})^2}}
\right]
d\phi(\mathbf{k}),
\label{eq:Phi_Delta_rho}
\end{equation}
where $d\phi(\mathbf{k})\equiv \nabla_{\mathbf{k}}\phi(\mathbf{k})\cdot d\mathbf{k}$.

Finally, expressing $\Delta$ and $\rho$ back in terms of the original Hamiltonian parameters,
\begin{equation}
\Delta(\mathbf{k})=\frac{\epsilon_1(\mathbf{k})-\epsilon_2(\mathbf{k})}{2},
\qquad
\rho(\mathbf{k})=|g(\mathbf{k})|,
\end{equation}
we arrive at the general result
\begin{equation}
\boxed{
\Phi_\pm[C]
=
\pm \frac{1}{2}
\oint_C
\left[
1-
\frac{\epsilon_1(\mathbf{k})-\epsilon_2(\mathbf{k})}
{\sqrt{\bigl(\epsilon_1(\mathbf{k})-\epsilon_2(\mathbf{k})\bigr)^2+4|g(\mathbf{k})|^2}}
\right]
d\,\arg g(\mathbf{k})
}.
\label{eq:Phi_general_eps12}
\end{equation}
This is the desired closed-form expression for the Berry phase of a generic two-band model with unequal diagonal dispersions.

\begin{figure}[t]
    \centering
    \begin{overpic}[width=0.3\linewidth]{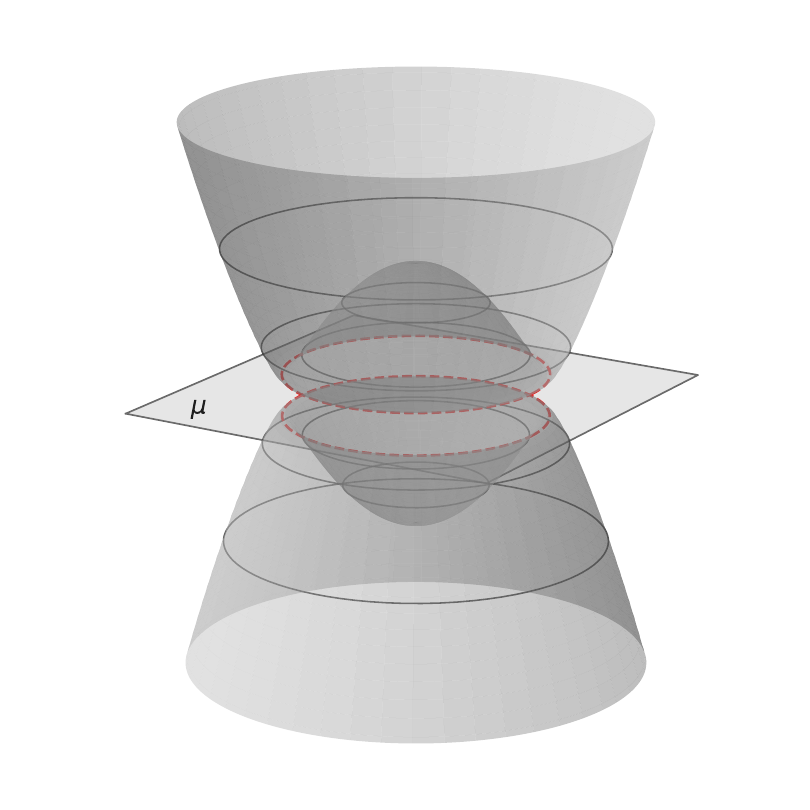}
    \end{overpic}
    \caption{
        Schematic of contour orbits for the Berry phase in momentum space. }
    \label{fig:Phi_k}
\end{figure}

For many applications, one is interested in an orbit of fixed radius $k$ in an isotropic model. Suppose that along such an orbit like in Fig.~\ref{fig:Phi_k}, the hybridization takes the form
\begin{equation}
g(\mathbf{k})=|g(k)|e^{-iw\theta},
\label{eq:g_isotropic}
\end{equation}
where $\theta$ is the polar angle in momentum space and $w$ is the winding number of the hybridization phase, which can be identified with the angular momentum in the maintext.
If, in addition, $\epsilon_1(k)$, $\epsilon_2(k)$, and $|g(k)|$ depend only on $k$, then $d\,\arg g(\mathbf{k})=-w\,d\theta$, and Eq.~\eqref{eq:Phi_general_eps12} reduces to
\begin{equation}
\Phi_\pm(k)
=
\mp \pi w
\left[
1-
\frac{\epsilon_1(k)-\epsilon_2(k)}
{\sqrt{\bigl(\epsilon_1(k)-\epsilon_2(k)\bigr)^2+4|g(k)|^2}}
\right].
\label{eq:Phi_isotropic_eps12}
\end{equation}
This form is particularly useful because it makes the physical content transparent: the Berry phase is determined by the competition between the diagonal energy difference $\epsilon_1-\epsilon_2$ and the off-diagonal mixing $g$.

For a fixed path $C$, Eq.~\eqref{eq:Phi_general_eps12} shows that the Berry phase is insensitive to the average energy $(\epsilon_1+\epsilon_2)/2$ and depends only on the difference $\epsilon_1-\epsilon_2$. Physically, this difference acts as an effective mass term for the pseudospin texture. When
\begin{equation}
|\epsilon_1(\mathbf{k})-\epsilon_2(\mathbf{k})|\gg 2|g(\mathbf{k})|,
\label{eq:large_mass_limit}
\end{equation}
the pseudospin points predominantly along the north or south pole of the Bloch sphere, and the Berry phase is close to zero. By contrast, when the diagonal dispersions become nearly degenerate along the relevant orbit,
\begin{equation}
\epsilon_1(\mathbf{k})\approx \epsilon_2(\mathbf{k}),
\label{eq:nearly_degenerate}
\end{equation}
the pseudospin lies close to the equator, and the Berry phase becomes maximally nontrivial. In particular, if there exists a closed orbit on which
\begin{equation}
\epsilon_1(\mathbf{k})=\epsilon_2(\mathbf{k}),
\label{eq:equal_eps_on_orbit}
\end{equation}
then Eq.~\eqref{eq:Phi_isotropic_eps12} gives
\begin{equation}
\Phi_\pm=\mp \pi w
\qquad (\mathrm{mod}\ 2\pi).
\label{eq:piw_limit}
\end{equation}
For $w=1$, this yields the familiar $\pi$ Berry phase.

Therefore, the case with unequal diagonal dispersions is a straightforward generalization of the symmetric two-band model: the essential control parameter is not whether the two diagonal terms are globally equal, but whether their difference is small or vanishes on the closed orbit relevant for the semiclassical motion.

\section{Semiclassical description of insulating-regime LEDOS oscillations}
\label{app:semiclassical_ledos}

In this Appendix, we organize the semiclassical derivation in the same sequence as the main text. We first isolate the oscillatory response of a single chiral two-band block, then formulate its energy-resolved Onsager quantization and thermally broadened LEDOS, and finally combine the two opposite chiralities into the time-reversal-invariant four-band response.

\vskip 0.5cm

We begin with a single chiral two-band block
\begin{equation}
H_{\eta}(\mathbf{k})
=
\begin{pmatrix}
\epsilon_1(k) & g_\eta(\mathbf{k}) \\
g_\eta^*(\mathbf{k}) & \epsilon_2(k)
\end{pmatrix},
\qquad
\eta=\pm ,
\label{eq:app_two_band_block}
\end{equation}
where \(\eta=\pm\) labels the two opposite chiralities that are later combined into the four-band model. We define
\begin{equation}
\epsilon_0(k)=\frac{\epsilon_1(k)+\epsilon_2(k)}{2},
\qquad
M(k)=\frac{\epsilon_1(k)-\epsilon_2(k)}{2}.
\end{equation}
For a high-angular-momentum band inversion, the hybridization
takes the chiral form
\begin{equation}
g_\eta(\mathbf{k})
=
V k^l e^{-i\eta l\theta_{\mathbf{k}}},
\label{eq:app_chiral_hybridization}
\end{equation}
where \(l\) is the relative angular momentum of the two inverted
bands. The two hybridized band energies are
\begin{equation}
E_{\lambda}(k)
=
\epsilon_0(k)
+
\lambda R(k),
\qquad
R(k)=\sqrt{M^2(k)+V^2k^{2l}},
\qquad
\lambda=\pm .
\label{eq:app_band_energy}
\end{equation}
Applying the general result of Sec.~\ref{app:berry_phase_eps1_eps2}
[Eq.~\eqref{eq:Phi_isotropic_eps12}] with hybridization winding number
\(w=\eta l\), the Berry phase of band \(\lambda\) in block \(\eta\) on
a circular constant-energy orbit of radius \(k\) is
\begin{equation}
\Phi_{\lambda,\eta}(k)
=
-\lambda \eta \pi l
\left[
1-\frac{M(k)}{R(k)}
\right],
\label{eq:app_berry_phase}
\end{equation}
up to a convention-dependent integer multiple of \(2\pi\).
This dependence on both \(\lambda\) and \(\eta\) has a simple origin.
The band index \(\lambda=\pm\) distinguishes the upper and lower
hybridized bands, whose pseudospins point in opposite directions on the
Bloch sphere and therefore acquire opposite Berry phases on the same
orbit. By contrast, the block index \(\eta=\pm\) labels the two
time-reversal-related chiralities of the hybridization,
\(g_\eta(\mathbf{k})\propto e^{-i\eta l\theta_{\mathbf{k}}}\), so it
controls the sense in which the pseudospin texture winds as the orbit is
traversed. Reversing either the band index or the block chirality flips
the sign of the Berry phase, which is why Eq.~\eqref{eq:app_berry_phase}
contains the product \(-\lambda\eta\).
The key property for the later four-band construction is the time-reversal relation
\begin{equation}
\Phi_{\lambda,+}(k)
=
-\Phi_{\lambda,-}(k)
\quad
\mathrm{mod}\;2\pi .
\label{eq:app_TR_berry_relation}
\end{equation}
At the band-inversion orbit, defined approximately by \(M(k_F)=0\),
one obtains
\begin{equation}
\Phi_{\lambda,\eta}(k_F)
=
-\lambda \eta \pi l .
\end{equation}
Thus the unwrapped Berry phase accumulated around the orbit is
controlled by the angular momentum \(l\).

\vskip 0.5cm

The semiclassical quantization condition for a closed orbit of band
\(\lambda\) at energy \(E\) is
\begin{equation}
\ell_B^2 S_\lambda(E)
=
2\pi
\left[
n+\gamma_{\lambda,\eta}(E)
\right],
\label{eq:app_onsager}
\end{equation}
where
\begin{equation}
\ell_B=\sqrt{\frac{\hbar}{eB}},
\qquad
S_\lambda(E)
=
\mathrm{area}
\left\{
\mathbf{k}: E_\lambda(\mathbf{k})=E
\right\},
\end{equation}
and
\begin{equation}
\gamma_{\lambda,\eta}(E)
=
\frac12
-
\frac{\Phi_{\lambda,\eta}(E)}{2\pi}.
\label{eq:app_gamma}
\end{equation}
Here \(\Phi_{\lambda,\eta}(E)\) means the Berry phase evaluated on
the constant-energy orbit \(E_\lambda(k)=E\).

It is useful to define the oscillation frequency
\begin{equation}
F_\lambda(E)
=
\frac{\hbar}{2\pi e}S_\lambda(E).
\label{eq:app_frequency}
\end{equation}
Then Eq.~\eqref{eq:app_onsager} becomes
\begin{equation}
\frac{F_\lambda(E)}{B}
=
n+\gamma_{\lambda,\eta}(E).
\label{eq:app_onsager_frequency}
\end{equation}
This is an energy-resolved quantization rule. In a metal, one
usually sets \(E=\mu\), so \(S_\lambda(\mu)\) is the Fermi-surface
area. In the insulating regime considered here, \(\mu\) lies in the
gap, but Eq.~\eqref{eq:app_onsager_frequency} still quantizes the
band-edge orbits at energies \(E\neq \mu\).

\vskip 0.5cm

The density of states per unit area for band \(\lambda\) in block
\(\eta\) can be written as
\begin{equation}
\rho_{\lambda,\eta}(E,B)
=
\frac{eB}{2\pi\hbar}
\sum_{n}
\delta\!\left(E-E_{\lambda,n,\eta}(B)\right).
\label{eq:app_DOS_LL}
\end{equation}
Using the quantization condition
\[
n =
\frac{F_\lambda(E)}{B}
-
\gamma_{\lambda,\eta}(E),
\]
we define
\[
x_{\lambda,\eta}(E)
=
\frac{F_\lambda(E)}{B}
-
\gamma_{\lambda,\eta}(E),
\]
so that
\[
\delta\!\left(E-E_{\lambda,n,\eta}(B)\right)
=
\left|\frac{\partial x_{\lambda,\eta}(E)}{\partial E}\right|
\delta\!\left(n-x_{\lambda,\eta}(E)\right).
\]
The Landau-level sum then becomes
\begin{equation}
\rho_{\lambda,\eta}(E,B)
=
\widetilde{\rho}_{\lambda,\eta}(E,B)
\sum_n
\delta
\left[
n
-
\frac{F_\lambda(E)}{B}
+
\gamma_{\lambda,\eta}(E)
\right],
\end{equation}
with
\begin{equation}
\widetilde{\rho}_{\lambda,\eta}(E,B)
=
\frac{eB}{2\pi\hbar}
\left|
\frac{\partial}{\partial E}
\left[
\frac{F_\lambda(E)}{B}
-
\gamma_{\lambda,\eta}(E)
\right]
\right|.
\end{equation}
In the weak-field semiclassical regime, the derivative of
\(F_\lambda(E)/B\) with respect to energy is enhanced by the explicit
factor \(1/B\), whereas \(\gamma_{\lambda,\eta}(E)\) varies only on the
underlying band-energy scale. We therefore neglect
\(\partial_E\gamma_{\lambda,\eta}(E)\) in this slowly varying prefactor. Using
\(F_\lambda(E)=\hbar S_\lambda(E)/(2\pi e)\), one then obtains
\(\widetilde{\rho}_{\lambda,\eta}(E,B)\approx \frac{e}{2\pi\hbar}\abs{\partial_E F_\lambda(E)}=\frac{1}{4\pi^2}\abs{\partial_E S_\lambda(E)}\equiv \rho^0_\lambda(E)\), namely the smooth zero-field density of states of band \(\lambda\). In the insulating LEDOS problem below, this smooth DOS is sampled mainly within the thermal window around the band edges. This approximation only affects the prefactor, and the Berry-phase contribution to the oscillation phase is kept below.
\begin{equation}
\rho^0_\lambda(E)
=
\frac{1}{4\pi^2}
\left|
\frac{\partial S_\lambda(E)}{\partial E}
\right|.
\end{equation}
Applying the Poisson summation formula,
\begin{equation}
\sum_n \delta(n-x)
=
\sum_{p=-\infty}^{\infty} e^{2\pi i p x},
\end{equation}
we obtain
\begin{equation}
\rho_{\lambda,\eta}(E,B)
=
\rho^0_\lambda(E)
\left\{
1
+
2\sum_{p=1}^{\infty}
R_p(E,B)
\cos
\left[
2\pi p
\left(
\frac{F_\lambda(E)}{B}
-
\gamma_{\lambda,\eta}(E)
\right)
\right]
\right\}.
\label{eq:app_DOS_osc_general}
\end{equation}
Here \(R_p(E,B)\) is a phenomenological damping factor, which may
include disorder broadening, finite lifetime, or magnetic-breakdown
effects. Keeping only the leading harmonic \(p=1\), we find
\begin{equation}
\rho^{\mathrm{osc}}_{\lambda,\eta}(E,B)
\simeq
2\rho^0_\lambda(E)R_1(E,B)
\cos
\left[
2\pi\frac{F_\lambda(E)}{B}
-\pi
+
\Phi_{\lambda,\eta}(E)
\right].
\label{eq:app_DOS_osc_leading}
\end{equation}
This equation shows explicitly that the Berry phase enters as a
phase offset of the quantum oscillation.

\vskip 0.5cm

The LEDOS measured near the chemical potential is a thermally
broadened density of states,
\begin{equation}
D(B,T)
=
\int dE\,
K_T(E-\mu)
\rho(E,B),
\label{eq:app_ledos_definition}
\end{equation}
where
\begin{equation}
K_T(E-\mu)
=
-\frac{\partial n_F(E-\mu)}{\partial E}
=
\frac{1}{4T}
\operatorname{sech}^2
\left(
\frac{E-\mu}{2T}
\right).
\label{eq:app_thermal_kernel}
\end{equation}
In the insulating regime, \(\mu\) lies inside the hybridization gap.
Therefore, at \(T=0\) and in the absence of lifetime broadening,
\(D(B,0)\) vanishes. At finite temperature, however,
\(K_T(E-\mu)\) samples band-edge states within an energy window
of order \(T\). Since these band-edge states are Landau quantized,
their oscillatory density of states contributes to the LEDOS.

Combining Eqs.~\eqref{eq:app_DOS_osc_leading} and
\eqref{eq:app_ledos_definition}, the oscillatory part of the LEDOS
from block \(\eta\) is
\begin{equation}
D^{\mathrm{osc}}_{\eta}(B,T)
=
\sum_{\lambda=\pm}
\int dE\,
K_T(E-\mu)
\rho^{\mathrm{osc}}_{\lambda,\eta}(E,B).
\label{eq:app_ledos_osc_integral}
\end{equation}
Substituting Eq.~\eqref{eq:app_DOS_osc_leading}, we obtain
\begin{equation}
D^{\mathrm{osc}}_{\eta}(B,T)
\simeq
2
\sum_{\lambda}
\int dE\,
K_T(E-\mu)
\rho^0_\lambda(E)
R_1(E,B)
\cos
\left[
2\pi\frac{F_\lambda(E)}{B}
-\pi
+
\Phi_{\lambda,\eta}(E)
\right].
\label{eq:app_ledos_osc_full}
\end{equation}

Equation~\eqref{eq:app_ledos_osc_full} is the central expression
for the insulating-regime quantum oscillation. It differs from the
standard metallic Lifshitz-Kosevich formula in an essential way:
the oscillation is not determined by the Berry phase at a Fermi
surface. Instead, it is determined by a thermal average over
band-edge orbits.

\vskip 0.5cm

In the small-gap insulating regime, the dominant contribution to
Eq.~\eqref{eq:app_ledos_osc_full} comes from the thermally
activated band-edge states. If the oscillation frequency varies weakly
within the relevant thermal window, we may write
\begin{equation}
F_\lambda(E)
=
F_0+\delta F_\lambda(E),
\qquad
|\delta F_\lambda(E)|\ll F_0.
\end{equation}
Then Eq.~\eqref{eq:app_ledos_osc_full} can be organized as
\begin{equation}
D^{\mathrm{osc}}_{\eta}(B,T)
\simeq
2\operatorname{Re}
\left[
e^{i2\pi F_0/B}
Z_\eta(B,T)
\right],
\label{eq:app_ledos_complex}
\end{equation}
where
\begin{equation}
Z_\eta(B,T)
=
\sum_{\lambda}
\int dE\,
K_T(E-\mu)
\rho^0_\lambda(E)
R_1(E,B)
\exp
\left\{
i
\left[
2\pi\frac{\delta F_\lambda(E)}{B}
-\pi
+
\Phi_{\lambda,\eta}(E)
\right]
\right\}.
\label{eq:app_complex_amplitude_general}
\end{equation}
If the frequency dispersion \(\delta F_\lambda(E)\) is negligible
over the thermal window, \(Z_\eta(B,T)\) becomes approximately
a function of temperature only:
\begin{equation}
Z_\eta(T)
\simeq
\sum_{\lambda}
\int dE\,
K_T(E-\mu)
\rho^0_\lambda(E)
R_1(E)
e^{i[-\pi+\Phi_{\lambda,\eta}(E)]}.
\label{eq:app_complex_amplitude}
\end{equation}
Writing
\begin{equation}
Z_\eta(T)
=
A_\eta(T)e^{i\phi_\eta(T)},
\label{eq:app_A_phi_definition}
\end{equation}
we arrive at
\begin{equation}
D^{\mathrm{osc}}_{\eta}(B,T)
\simeq
2A_\eta(T)
\cos
\left[
2\pi\frac{F_0}{B}
+
\phi_\eta(T)
\right].
\label{eq:app_single_block_final}
\end{equation}

The phase \(\phi_\eta(T)\) is therefore not a constant Berry phase.
Rather, it is the phase of a thermally averaged complex amplitude.
As temperature increases, the thermal kernel samples a wider range
of band-edge orbits, and the effective Berry phase changes
accordingly.

\vskip 0.5cm

We next combine these two opposite chiralities into the time-reversal-invariant four-band model, which consists of two blocks with opposite chirality:
\begin{equation}
H_{\mathrm{4band}}(\mathbf{k})
=
\begin{pmatrix}
H_+(\mathbf{k}) & 0 \\
0 & H_-(\mathbf{k})
\end{pmatrix},
\qquad
H_-(\mathbf{k})
=
\mathcal{T}H_+(-\mathbf{k})\mathcal{T}^{-1}.
\end{equation}
Because the two blocks are related by time reversal, their
cyclotron areas and smooth densities of states are the same, while
their Berry phases have opposite signs:
\begin{equation}
F_{\lambda,+}(E)=F_{\lambda,-}(E),
\qquad
\rho^0_{\lambda,+}(E)=\rho^0_{\lambda,-}(E),
\qquad
\Phi_{\lambda,+}(E)=-\Phi_{\lambda,-}(E).
\end{equation}
In the symmetric limit, Eq.~\eqref{eq:app_complex_amplitude}
therefore gives
\begin{equation}
Z_-(T)
=
Z_+^*(T).
\end{equation}
Equivalently,
\begin{equation}
A_+(T)=A_-(T)\equiv A(T),
\qquad
\phi_-(T)=-\phi_+(T).
\end{equation}
More generally, weak particle-hole asymmetry or weak block
asymmetry can make the two amplitudes slightly different, but the
two phases still shift in opposite directions because the Berry
phases are opposite.

The total oscillatory LEDOS is
\begin{equation}
D^{\mathrm{osc}}_{\mathrm{tot}}(B,T)
=
D^{\mathrm{osc}}_+(B,T)
+
D^{\mathrm{osc}}_-(B,T).
\end{equation}
Using Eq.~\eqref{eq:app_single_block_final}, one obtains
\begin{align}
D^{\mathrm{osc}}_{\mathrm{tot}}(B,T)
&\simeq
2A_+(T)
\cos
\left[
2\pi\frac{F_0}{B}
+
\phi_+(T)
\right]
+
2A_-(T)
\cos
\left[
2\pi\frac{F_0}{B}
+
\phi_-(T)
\right].
\end{align}
For approximately equal amplitudes,
\(A_+(T)\simeq A_-(T)\equiv A(T)\), this becomes
\begin{equation}
D^{\mathrm{osc}}_{\mathrm{tot}}(B,T)
\simeq
4A(T)
\cos
\left[
\frac{\Delta\phi(T)}{2}
\right]
\cos
\left[
2\pi\frac{F_0}{B}
+
\bar{\phi}(T)
\right],
\label{eq:app_interference_formula}
\end{equation}
where
\begin{equation}
\Delta\phi(T)=\phi_+(T)-\phi_-(T),
\qquad
\bar{\phi}(T)=\frac{\phi_+(T)+\phi_-(T)}{2}.
\end{equation}
Thus the observable amplitude of the leading oscillation is
\begin{equation}
A_{\mathrm{tot}}(T)
\simeq
4A(T)
\left|
\cos
\frac{\Delta\phi(T)}{2}
\right|.
\label{eq:app_total_amplitude}
\end{equation}
The leading oscillation is suppressed when
\begin{equation}
\Delta\phi(T_c)
=
(2j+1)\pi,
\qquad
j\in\mathbb{Z}.
\label{eq:app_destructive_condition}
\end{equation}
This is the destructive-interference condition between the two
time-reversal-related blocks.

If the two block amplitudes are not exactly equal, the total complex
amplitude is
\begin{equation}
Z_{\mathrm{tot}}(T)
=
A_+(T)e^{i\phi_+(T)}
+
A_-(T)e^{i\phi_-(T)},
\end{equation}
and hence
\begin{equation}
|Z_{\mathrm{tot}}(T)|
=
\sqrt{
A_+^2(T)
+
A_-^2(T)
+
2A_+(T)A_-(T)\cos\Delta\phi(T)
}.
\label{eq:app_unequal_amplitude}
\end{equation}
In this more general case, the destructive interference produces
a strong minimum rather than a perfect zero.

\vskip 0.5cm

The number of suppression temperatures is controlled by the
unwrapped phase evolution. Near the inversion orbit \(M(k_F)=0\),
the Berry phase is approximately
\begin{equation}
\Phi_{\lambda,\eta}(k_F)
=
-\lambda\eta \pi l .
\end{equation}
Away from the inversion orbit, where \(|M(k)|\gg Vk^l\), the Berry
phase becomes trivial modulo \(2\pi\). Therefore, as the thermally
sampled orbit changes with temperature, the effective phase
\(\phi_\eta(T)\) can sweep an \(l\)-dependent range. Since the two
time-reversal-related blocks have opposite chirality, their relative
phase can sweep approximately
\begin{equation}
\Delta\phi(T)
\sim
2l\pi
\end{equation}
in the unwrapped representation. The condition
\(\Delta\phi(T_c)=(2j+1)\pi\) can then naturally be satisfied up to
\(l\) times, provided that the phase evolves approximately
monotonically with temperature and the leading harmonic dominates.
In this sense, the semiclassical picture explains why the numerically
observed number of characteristic suppression temperatures tracks the
angular momentum \(l\), rather than constituting a strict counting
theorem.

\section{Landau levels and LEDOS oscillations}

For our calculations, we fix the chemical potential at $\mu=E_F$, take $m_d=1$ and $m_f=11$, choose $V=0.1, 0.05, 0.01, 0.002$ for $l=0,1,2,3$, and set $\hbar=e=1$ for simplicity. We set $\delta\mu=30$ unless otherwise specified.

\begin{figure}[h]
    \centering
    \begin{overpic}[width=0.4\columnwidth]{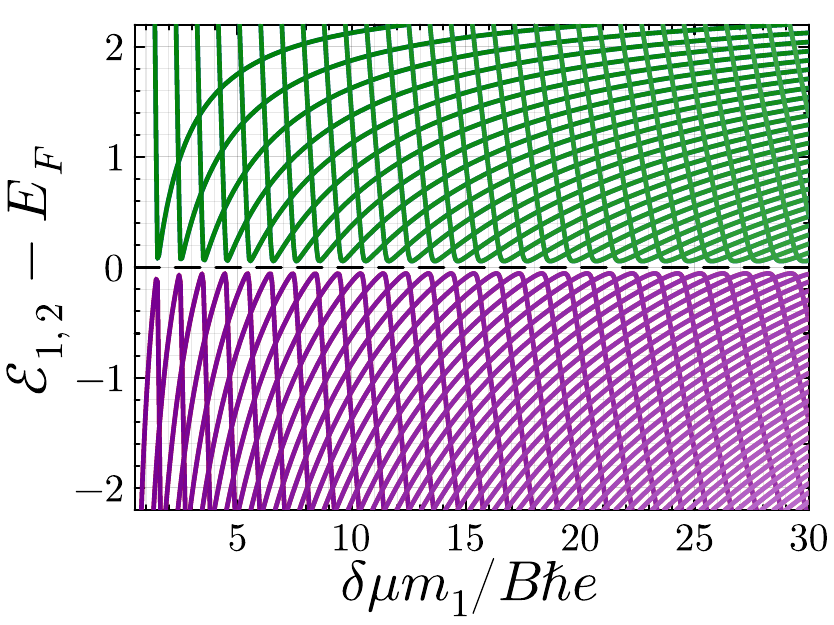}
        \put(0,70){(a)}
    \end{overpic}
    \begin{overpic}[width=0.4\columnwidth]{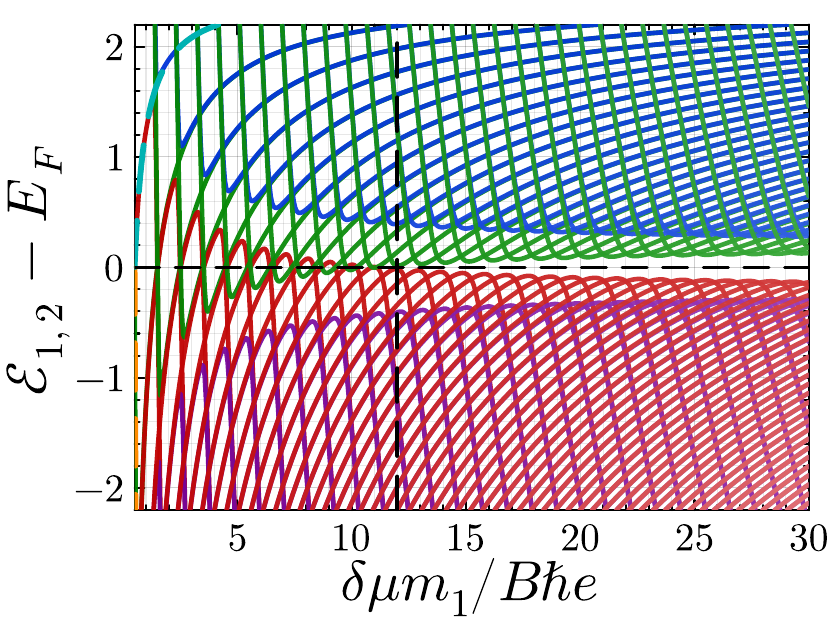}
        \put(0,70){(b)}
    \end{overpic}
    \begin{overpic}[width=0.4\columnwidth]{LanLvls_ll2_W30_ma1.0_mb11.0_v0.0100_vs_1overB_all.pdf}
        \put(0,70){(c)}
    \end{overpic}
    \begin{overpic}[width=0.4\columnwidth]{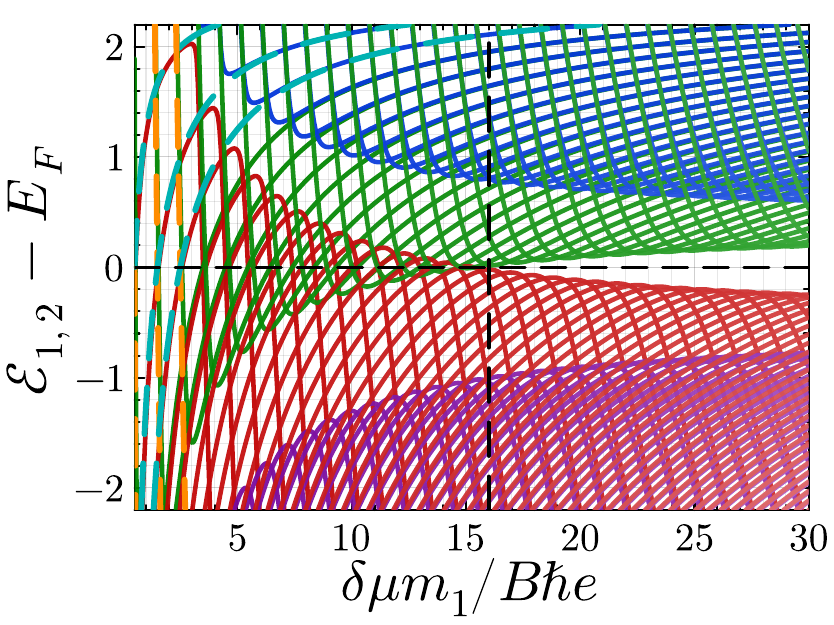}
        \put(0,70){(d)}
    \end{overpic}
    \caption{Landau level spectra $\mathcal{E}_{n}^{1,2}$ [Eq.~\eqref{eq:LLeigen}] plotted versus the inverse magnetic field $1/B$ (in units of $e\hbar/\delta\mu m_1$) for (a) $l=0$, (b) $l=1$, (c) $l=2$, and (d) $l=3$. The parameters are $m_d=1.0$, $m_f=11.0$, $\delta\mu=30.0$, with hybridization strengths $V=0.1, 0.05, 0.01, 0.002$ for $l=0,1,2,3$ respectively. The gray dashed line marks the critical field $B_c$ separating the insulating-like regime at low field from the metallic-like regime at high field. Note that larger angular momentum requires progressively weaker hybridization to open a comparable gap at the band-crossing point.}
    \label{fig:LL_l3l0}
\end{figure}

\begin{figure}[t]
    \begin{overpic}[width=0.4\linewidth]{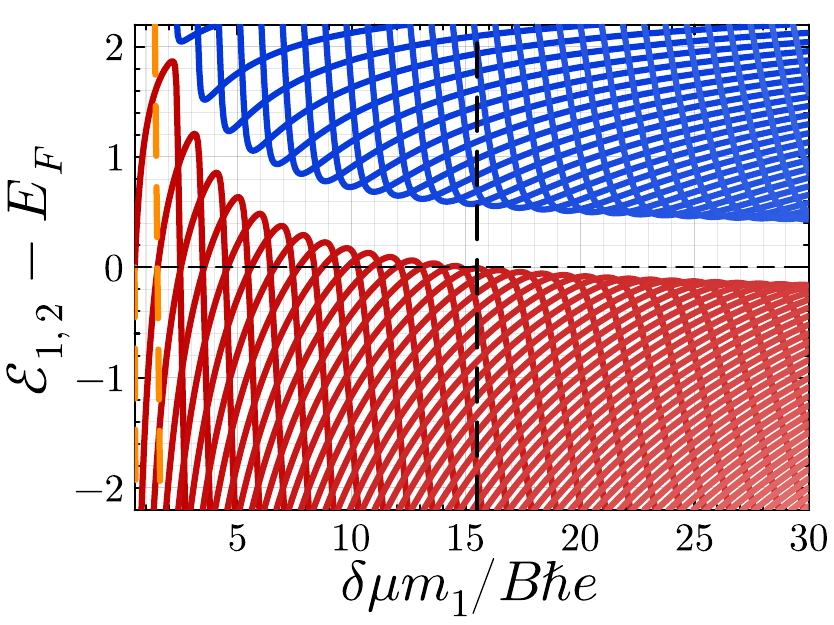}
        \put(0,72){(a1)} % 在左上角添加标号
    \end{overpic}
    \begin{overpic}
        [width=0.4\linewidth]{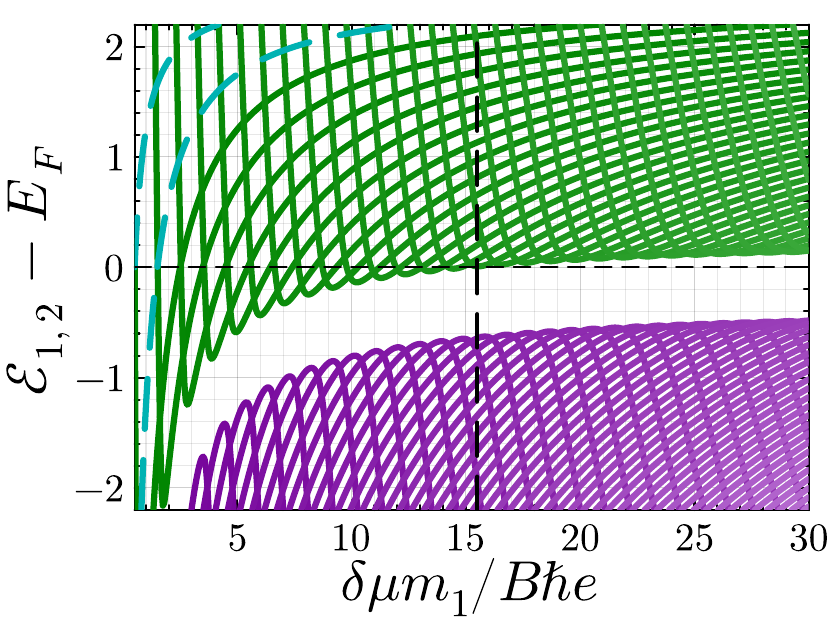}
        \put(0,72){(a2)} % 在左上角添加标号
    \end{overpic}
    \begin{overpic}[width=0.4\linewidth]{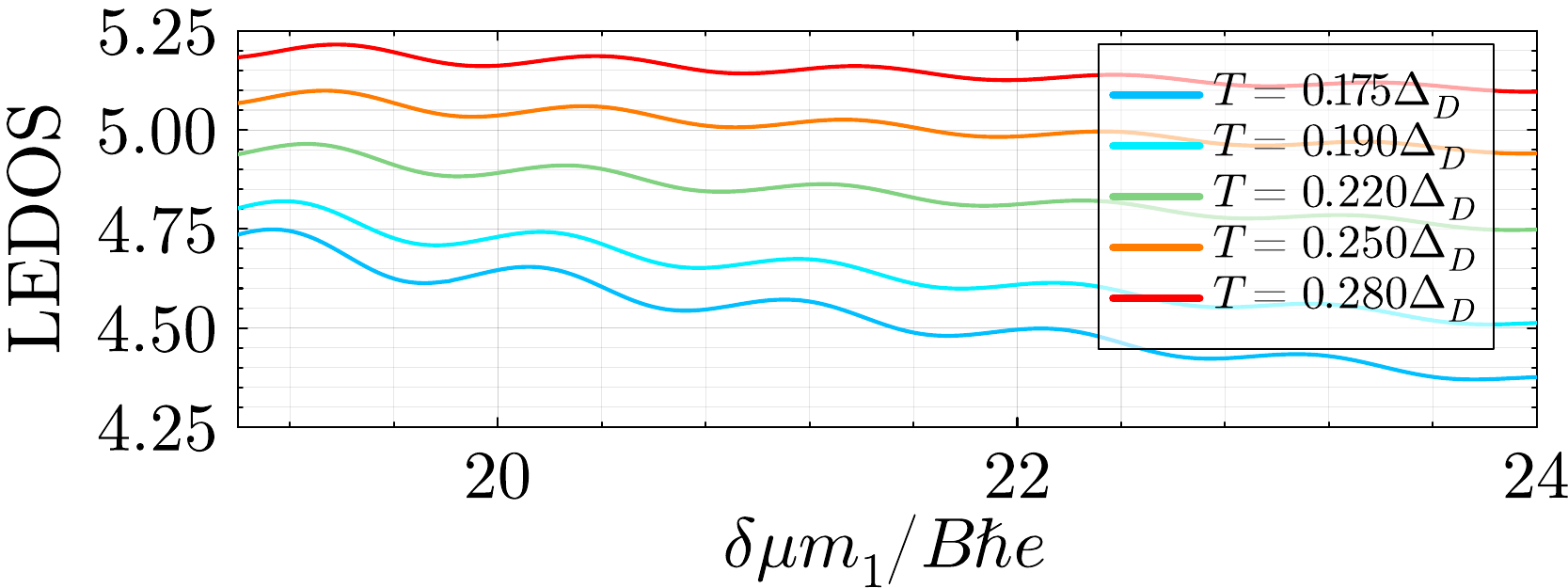}
        \put(0,52){(b1)} % 在左上角添加标号
    \end{overpic}
    \begin{overpic}[width=0.4\linewidth]{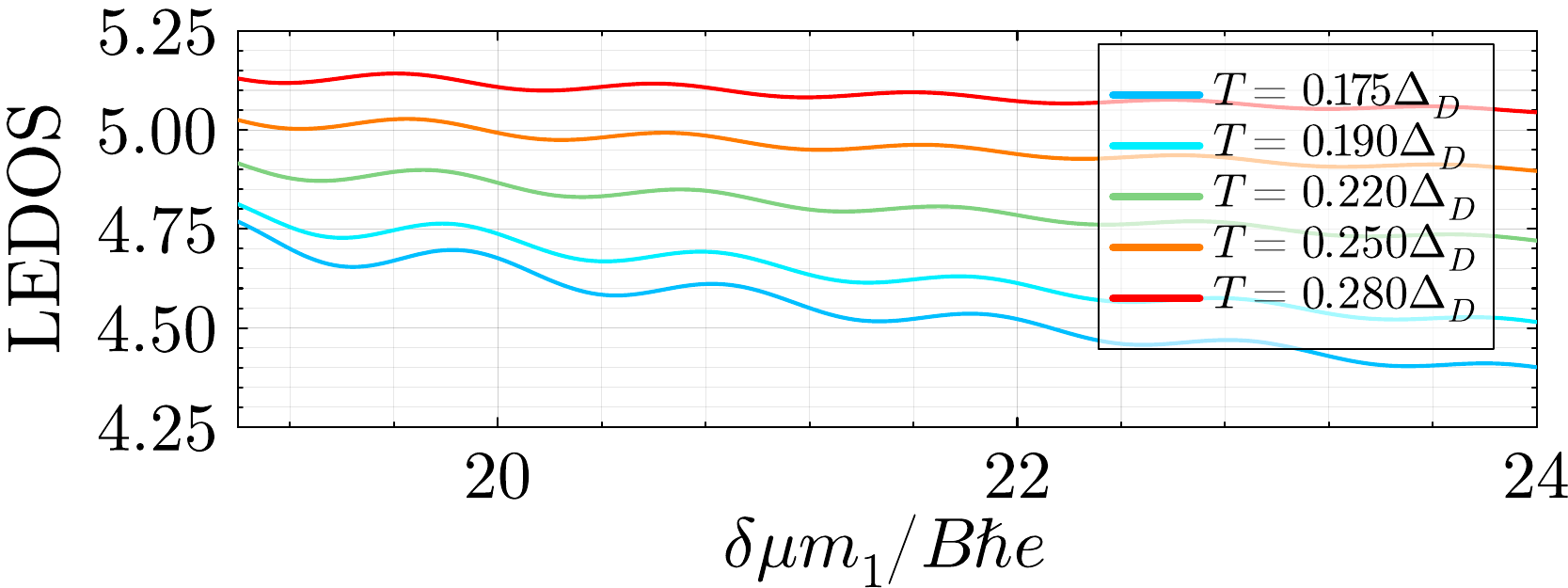}
        \put(0,52){(b2)} % 在左上角添加标号
    \end{overpic}
    \begin{overpic}
        [width=0.4\linewidth]{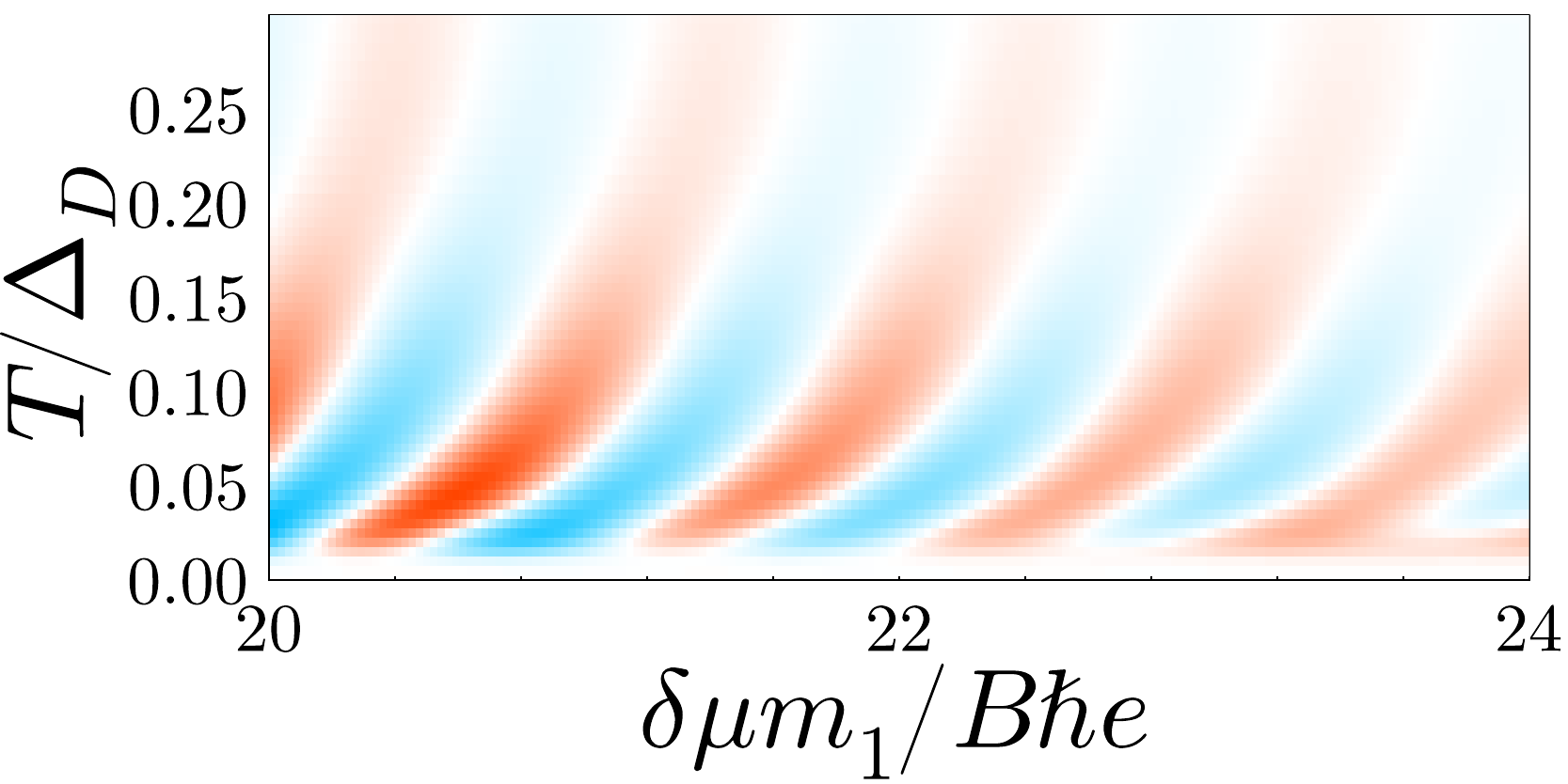}
        \put(0,51){(c1)} % 在左上角添加标号
    \end{overpic}
    \begin{overpic}
        [width=0.4\linewidth]{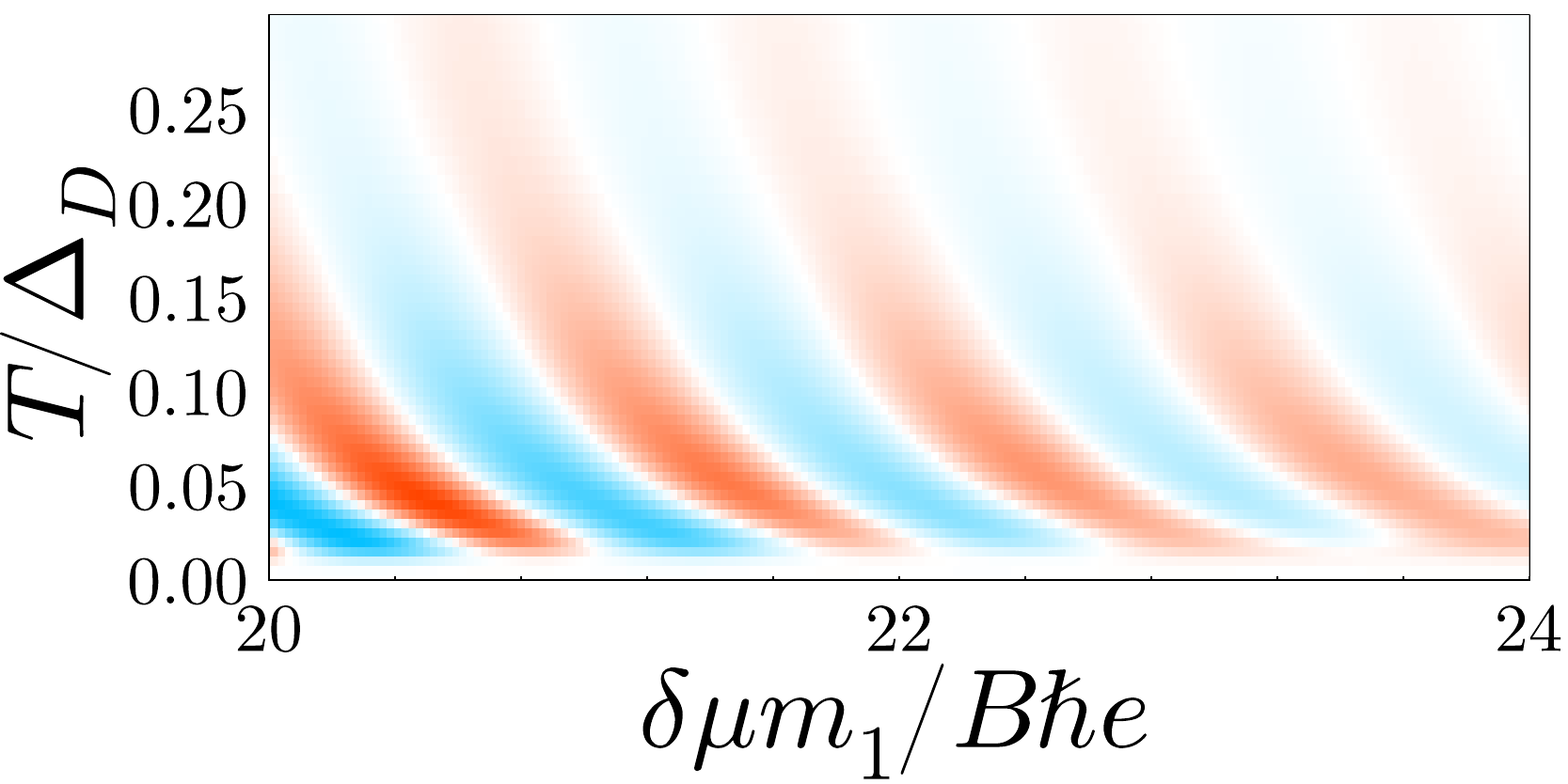}
        \put(0,51){(c2)} % 在左上角添加标号
    \end{overpic}
    \caption{
        (a1,a2) Landau levels $\mathcal{E}_n^{1,2}$ for $H_\uparrow\equiv \mathcal{H}_{\bm{k}}$ and $H_\downarrow\equiv\mathcal{H}_{-\bm{k}}^*$, respectively, plotted versus $\delta\mu m_1/B\hbar e$. 
        (b1,b2) Temperature dependence of the oscillation peaks for $H_\uparrow$ and $H_\downarrow$, respectively.  
        (c1,c2) The corresponding detrended LEDOS oscillations. The peaks in the two blocks shift in opposite directions with temperature because of their opposite chiral couplings. 
    }
    % \label{fig:Hp_Hm_LL_LEDOS}
\end{figure}

\begin{figure}[h]
    \centering
    \begin{overpic}[width=0.3\linewidth]{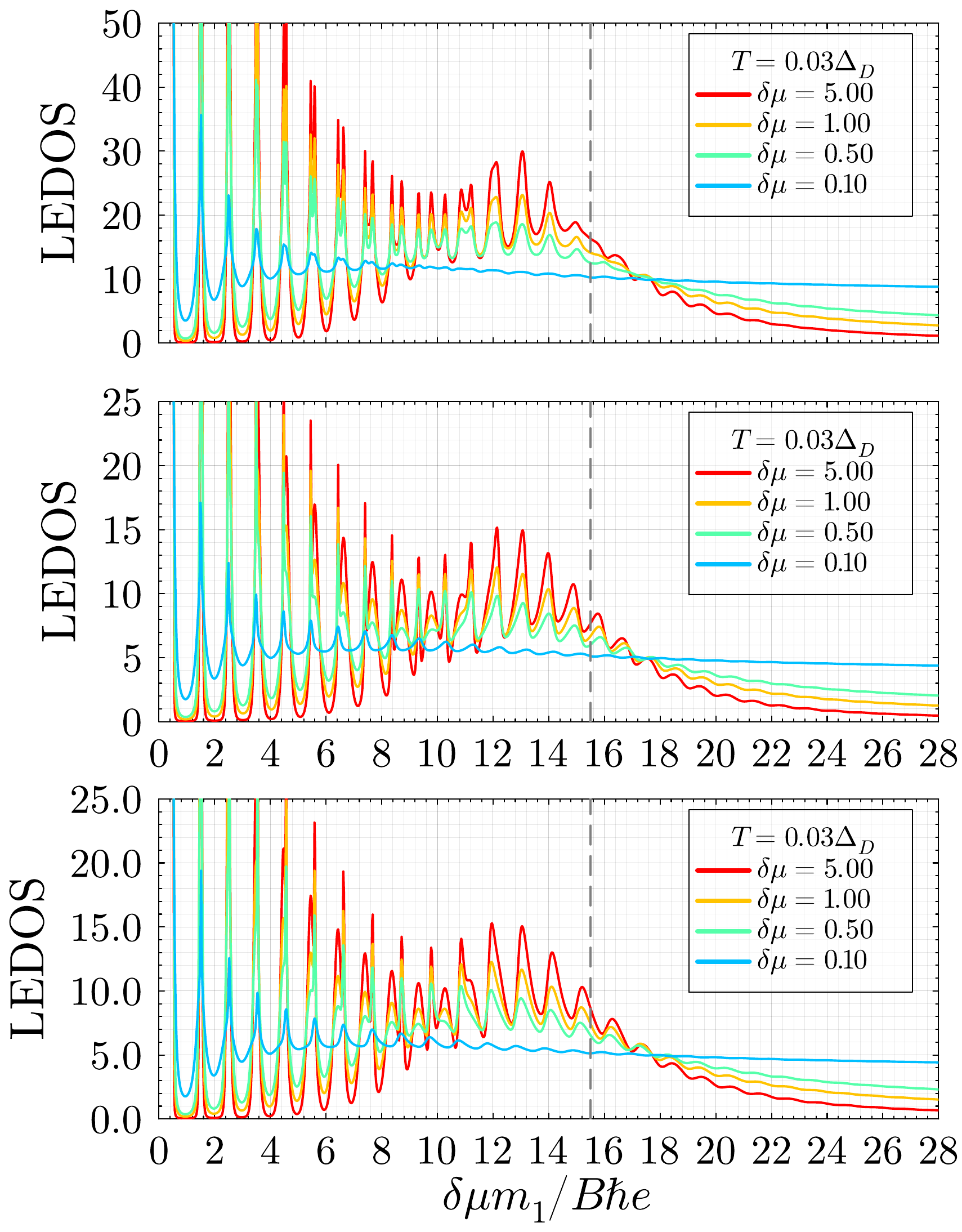}
        \put(-2,100){(a)} % 在左上角添加标号
    \end{overpic}
    \begin{overpic}
    [width=0.3\linewidth]{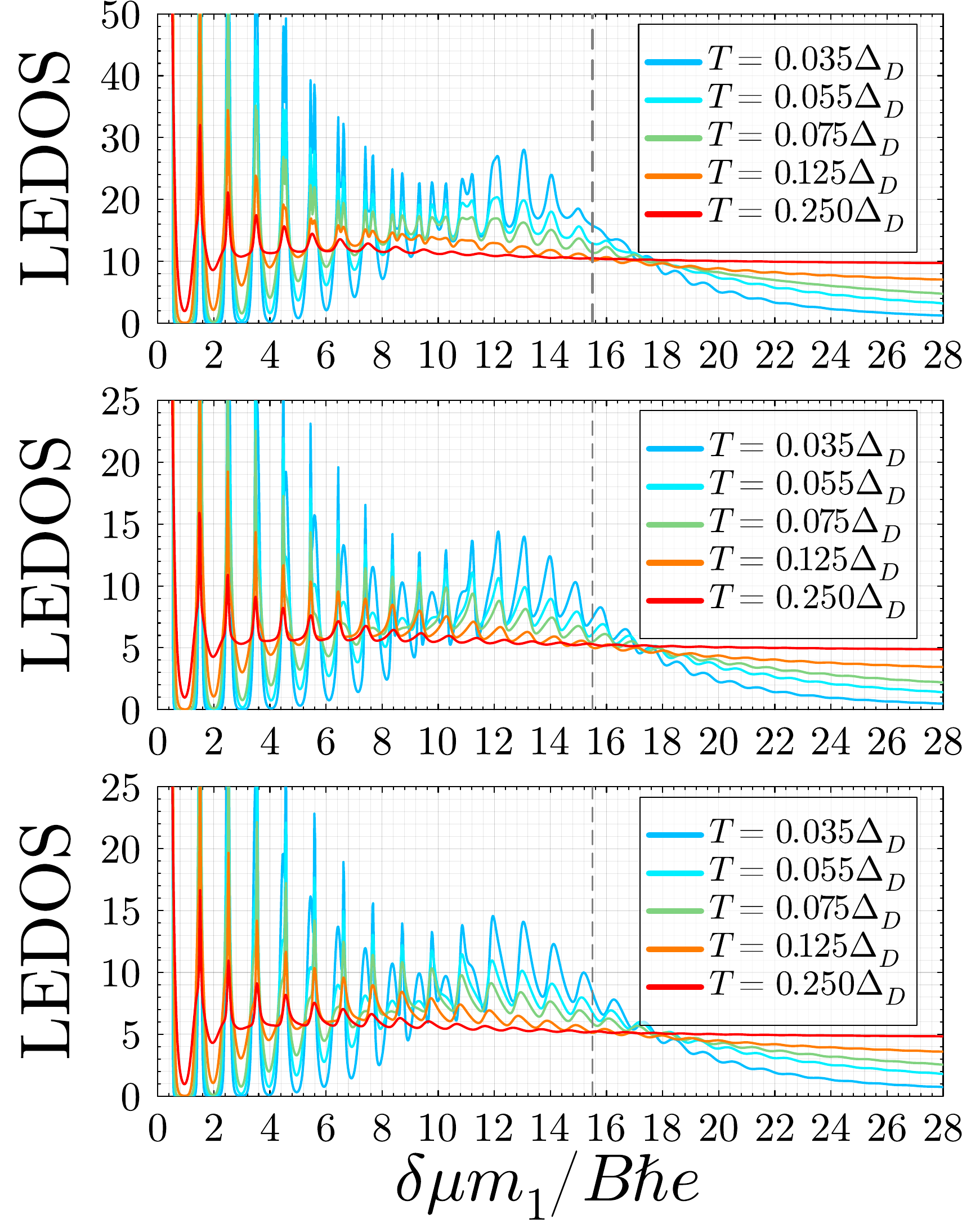}
    \put(-2,100){(b)} % 在左上角添加标号
    \end{overpic}
    % \\
    \begin{overpic}[width=0.3\linewidth]{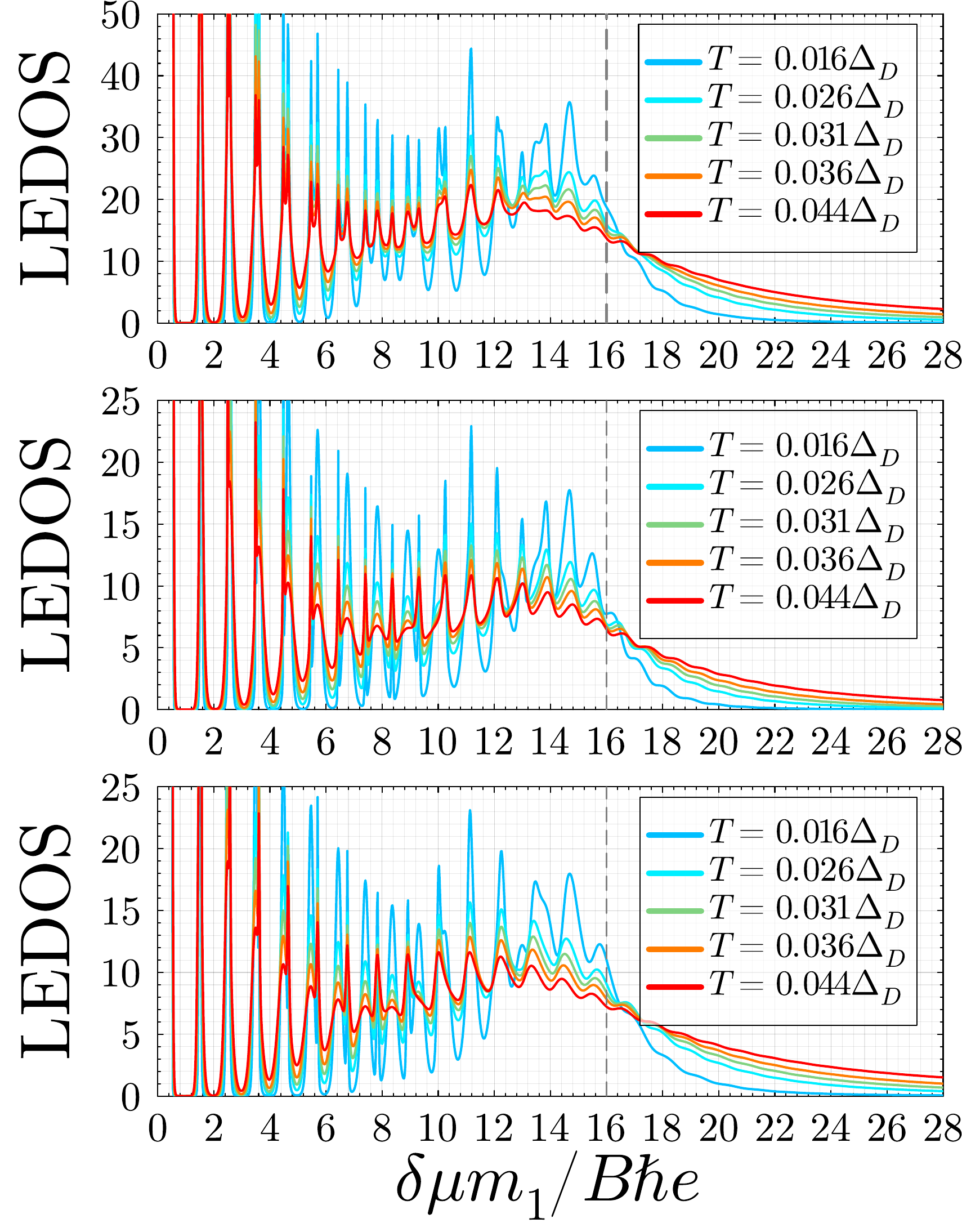}
        \put(-2,100){(c)} % 在左上角添加标号
    \end{overpic}
    \caption{LEDOS oscillations. (a) Comparison for different bandwidths $W$ at fixed angular momentum $l=2$. (b,c) Temperature dependence of the LEDOS oscillations in the insulating regime for $l=2$ and $l=3$, respectively.
    The two time-reversal-related blocks $H_\uparrow$ (middle) and $H_\downarrow$ (bottom), and their total superposition (top). }
\end{figure}

\begin{figure}[h]
    \centering
    \begin{overpic}
        [width=0.35\linewidth]{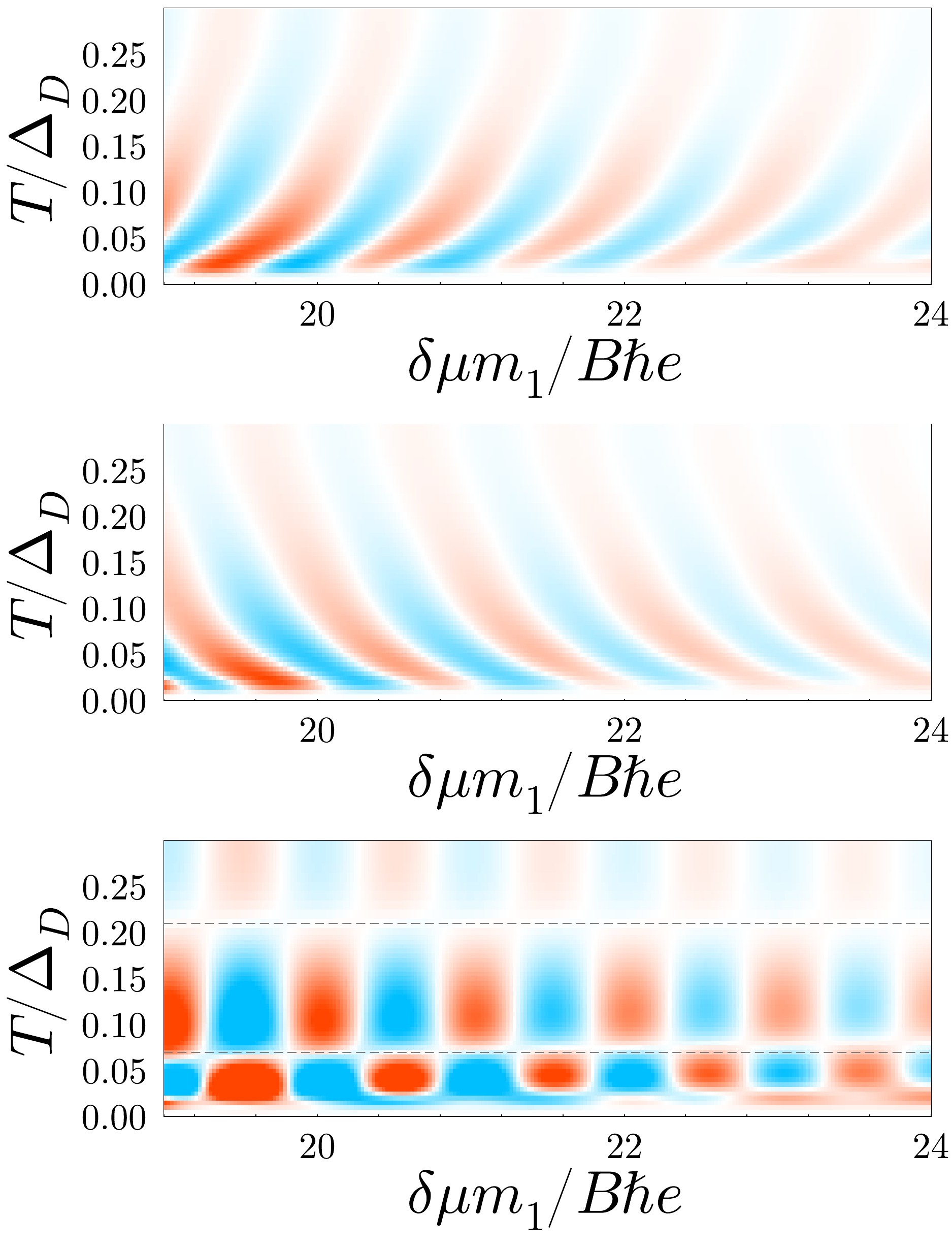}
        \put(0,95){(a1)} % 在左上角添加标号
        \put(0,65){(b1)} % 在左上角添加标号
        \put(0,33){(c1)} % 在左上角添加标号
    \end{overpic}
    \begin{overpic}
        [width=0.35\linewidth]
        {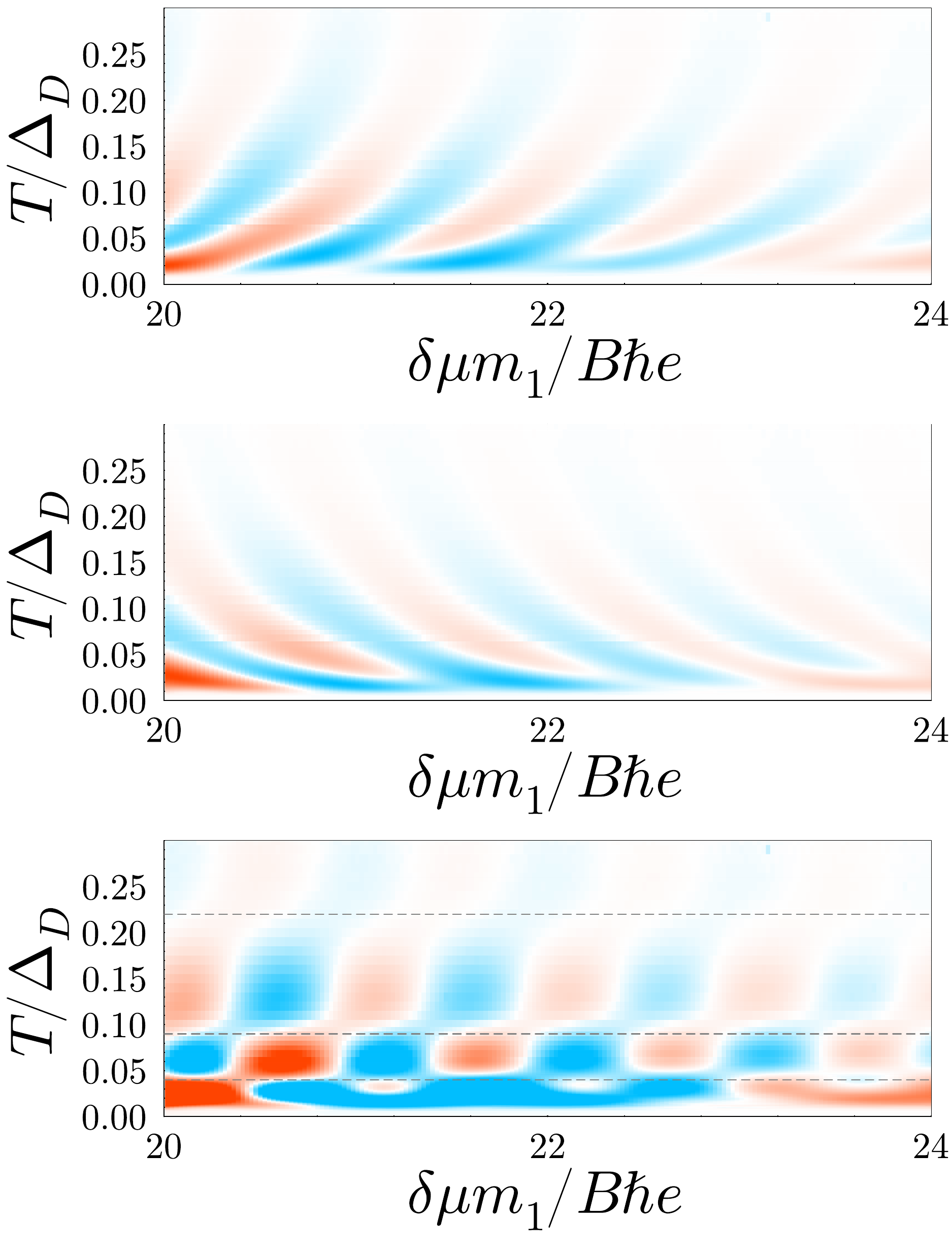}
        \put(0,95){(a2)} % 在左上角添加标号
        \put(0,65){(b2)} % 在左上角添加标号
        \put(0,33){(c2)} % 在左上角添加标号
    \end{overpic}
    \caption{Detrended LEDOS oscillations for (a1-a3) $l=2$, and (b1-b3) $l=3$, showing the two time-reversal-related blocks $H_\uparrow$ (top) and $H_\downarrow$ (middle), and their total superposition (bottom). The interference between the two blocks generates characteristic suppression temperatures whose number tracks $l$ in the examples shown.}
\end{figure}

\begin{figure}[t]
    % \vskip 0.5em
    \begin{overpic}[width=0.35\linewidth]{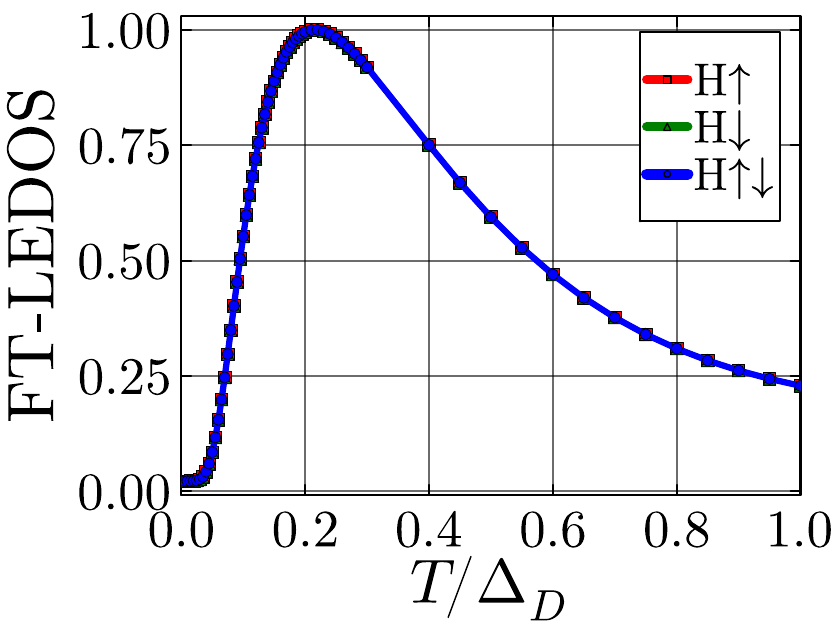}
        \put(-5,70){(a)} % 在左上角添加标号
    \end{overpic}
    \begin{overpic}[width=0.35\linewidth]{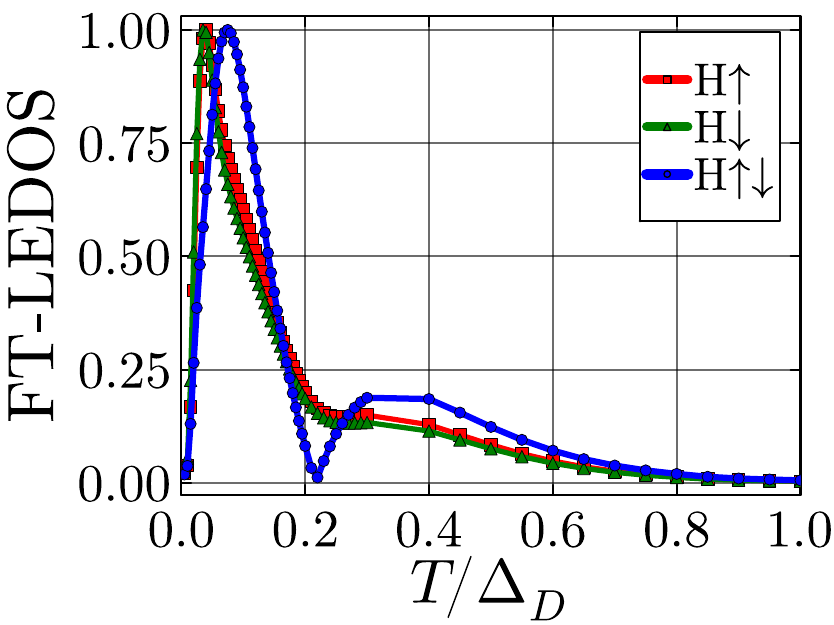}
        \put(-5,70){(b)} % 在左上角添加标号
    \end{overpic}
    \begin{overpic}[width=0.35\linewidth]{FFT_vs_T_freq1.00_AM2_W30.0_BinvMin19.00.pdf}
        \put(-5,70){(c)} % 在左上角添加标号
    \end{overpic}
    \begin{overpic}
    [width=0.35\linewidth]{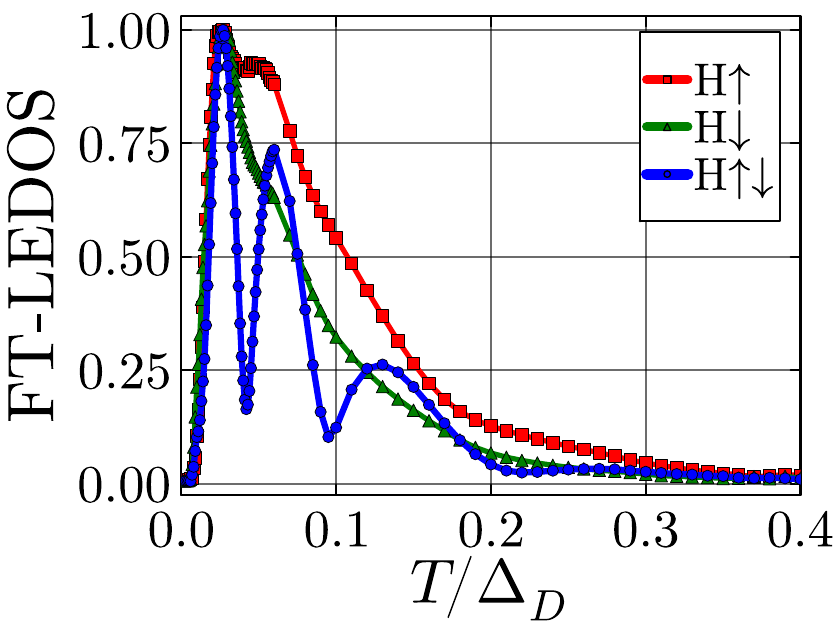}
        \put(-5,70){(d)} % 在左上角添加标号
    \end{overpic}
    \caption{
    Fourier spectra of the insulating-regime LEDOS for different angular momenta: (a) $l=0$, (b) $l=1$, (c) $l=2$, and (d) $l=3$. 
        In each panel, the spectrum is normalized by its maximum value. 
        The progressive suppression of spectral weight, accompanied here by characteristic temperatures $T_c$ whose number tracks $l$, reflects destructive interference between the two time-reversal counterparts. 
    }
    \label{fig:FT-LEDOS}
\end{figure}

\end{document}